\documentclass[%
    11pt,
    DIV=11,
    BCOR=0mm,
    twoside=semi,
    headings=small
]{scrartcl}

\usepackage{scrpage2}

\addtokomafont{title}{\normalfont \bfseries}
\addtokomafont{section}{\normalfont \bfseries}
\addtokomafont{subsection}{\normalfont \bfseries}
\addtokomafont{subsubsection}{\normalfont \bfseries}
\addtokomafont{paragraph}{\normalfont \bfseries}
\newcommand{\quoteparagraph}[1]{\noindent{\normalsize \bfseries #1}\hspace{1em}}

\usepackage[utf8]{inputenc}         % support for character encoding
\usepackage{amsmath, amsthm, amssymb} % pakages of American Mathematical Society (AMS)
\usepackage[english]{babel}           % support for English language
\usepackage{bibgerm}                  % support for german citation style
\usepackage{graphicx}                 % grafics (LaTeX: eps; pdfLaTeX: pdf, jpg, png)
\usepackage{hyperref}                 % automatic cross references
\usepackage{booktabs}                 % for nice tables
\usepackage{nicefrac}
\usepackage{subfig}
\usepackage{multirow}

\newcommand{\R}{\ensuremath{\mathbb{R}}}
\newcommand{\Z}{\ensuremath{\mathbb{Z}}}
\newcommand{\N}{\ensuremath{\mathbb{N}}}

\newtheorem{definition}{Definition}
\newtheorem{remark}{Remark}

\usepackage{bookmark}
\hypersetup{
  pdftitle    = {Empirical Study of the 1-2-3 Trend Indicator},
  pdfauthor   = {Yasemin Hafizogullari, Stanislaus Maier-Paape, and Andreas Platen},
  pdfkeywords = {market technical trends, automatic trend analysis, wavelength of time series, autocorrelation},
  pdfborder={0 0 0.5}
}

\author{
  \normalsize \textsc{Yasemin Hafizogullari}\\[-0.2em]
    \small \textit{Institut f\"ur Mathematik, RWTH Aachen,}\\[-0.5em]
    \small \textit{Templergraben 55, D-52052 Aachen, Germany}\\[-0.5em]
    \small \href{mailto:yasemin.hafizogullari@rwth-aachen.de}{yasemin.hafizogullari@rwth-aachen.de}\\
  \\[-0.75em]
  \normalsize \textsc{Stanislaus Maier-Paape}\\[-0.2em]
    \small \textit{Institut f\"ur Mathematik, RWTH Aachen,}\\[-0.5em]
    \small \textit{Templergraben 55, D-52052 Aachen, Germany}\\[-0.5em]
    \small \href{mailto:maier@instmath.rwth-aachen.de}{maier@instmath.rwth-aachen.de}\\
  \\[-0.75em]
  \normalsize \textsc{Andreas Platen}\\[-0.2em]
    \small \textit{Institut f\"ur Mathematik, RWTH Aachen,}\\[-0.5em]
    \small \textit{Templergraben 55, D-52052 Aachen, Germany}\\[-0.5em]
    \small \href{mailto:platen@instmath.rwth-aachen.de}{platen@instmath.rwth-aachen.de}
}
\date{
  \vspace{0.25em}
  \normalsize\today
  \vspace{-1cm}
}
\title{
  \vspace{-2cm}
  \Large Empirical Study of the 1--2--3 Trend Indicator
}

\clearscrheadfoot
\lehead{\pagemark}                   %% Top left on even pages
\rohead{\pagemark}                   %% Top right on odd pages
\begin{document}

\maketitle

\begin{quote}
  \small
  \quoteparagraph{Abstract}
  In this paper we study automatically recognized trends and investigate their statistics. To do that we introduce the notion
  of a wavelength for time series via cross correlation and use this wavelength to calibrate the 1--2--3 trend indicator of
  \cite{Maier-Paape2013} to automatically find trends. Extensive statistics are reported for EUR-USD, DAX-Future, Gold and Crude
  Oil regarding e.g. the dynamic, duration and extension of trends on different time scales.

  \quoteparagraph{Keywords}
  market technical trends,
  automatic trend analysis,
  wavelength of time series,
  autocorrelation
  
  \quoteparagraph{JEL classification}
  % C: Mathematical and quantitative methods
  %% C1: Econometric and Statistical Methods: General
  C19, % Other
  %% C6 Mathematical Methods; Programming Models; Mathematical and Simulation Modeling
  C63  % Computational techniques; Simulation modeling
\end{quote}

%%%%%%%%%%%%%%%%%%%% (1.)

\section{Introduction} % (1.)
\begin{figure} %1.1
  \centering
  \includegraphics[width=0.8\linewidth]{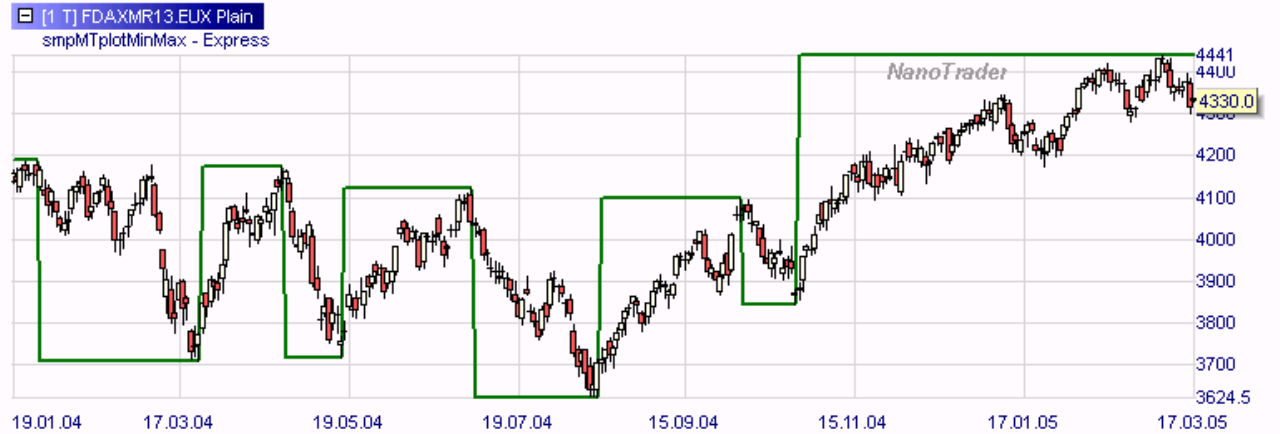}
  \caption{Snippet of the chart from DAX Future.}
  \label{fig:fdax}
\end{figure}

 If we take a look at an arbitrary financial time series, e.g. the chart of the DAX Future, it seems that the graph is affected by
 wavelike up and down movements, see Figure~\ref{fig:fdax}. Such movements are often used in e.g. trading systems or stopping criteria.
 This motivates us to study so called trends. Therefore we introduce the following notation of market mechanical up and down trends,
 basically going back to Charles H. Dow.

\begin{definition}\label{Trend_def} %1.1
  (Market mechanical up/down Trends)\\
  There is an up or down trend in a time series, if there is an increasing or a decreasing sequence of minima and maxima respectively,
  see Figure~\ref{fig:trend_definition}.
  \begin{itemize}
    \item Up trend: In this case the first extreme value must be a minimum, which we will call point~1. The second extreme value is a
    maximum (point~2) and the third one again a minimum (point 3). An up trend arises, if point~1 is below point~3 and the market price
    afterwards rises above the level of point~2, see Figure~\ref{fig:trend_definition}. The trend stays intact as long as the following
    extreme values are increasing minimal and maximal values.

    \item Down trend: This case is just a mirrored version of the up trend. It follows that point~1 is a maximum, point~2 a minimum and so on.
    As long as at least the last two minimal and last two maximal values are in a decreasing order, a down trend is active.
  \end{itemize}
\end{definition}

\begin{figure} %1.2
  \centering
  \includegraphics[width=0.35\linewidth]{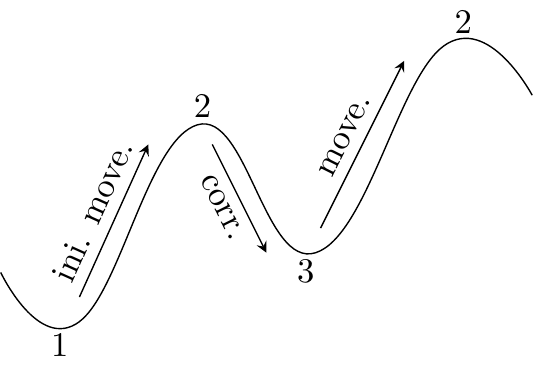}
  \caption{Up trend with numbering (cf. Definition~\ref{Trend_def}) and two movements.}
  \label{fig:trend_definition}
\end{figure}

The time periods between these extreme values have special names.

\begin{remark}\label{Bew_Korr} %1.
  The period of time between point 2 and 3 is called correction phase, whereas one says movement phase for the time
  from point 3 to the succeeding point~2.
\end{remark}

 Trends are a gateway between the psychological behavior of the traders and the market mechanics. For a heuristic introduction see \cite{Cene2011,Voigt2010}.
 Here we are interested in studying trends in a systematic way. For doing so it is important to understand the problematic of this definition of a trend or
 more specific the definition of a maximum and minimum.
 There are no rules to determine these minimal and maximal values. Therefore every arbitrary local extreme value can be used to identify trends, but clearly
 it is better when the extrema are ``significant''. This problem is demonstrated in Figure~\ref{fig:minmax_subjective}, where we can see one to at most three
 extreme values with continuously increasing significance. In any case the definition of relevant extrema is a very subjective issue.

\begin{figure} %1.3
  \centering
  \includegraphics[width=0.6\linewidth]{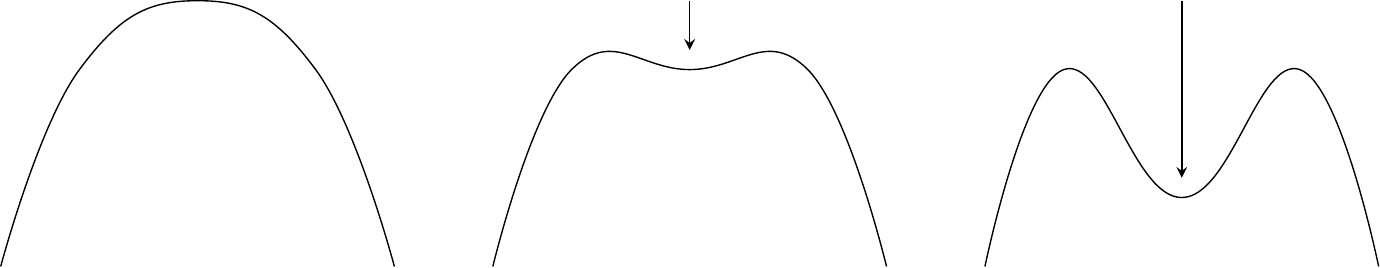}
  \caption{Continuation from one maximum to three extrema.}
  \label{fig:minmax_subjective}
\end{figure}

 In \cite{Maier-Paape2013} Maier-Paape describes a way to automatically generate a so called MinMax series out of a time series.
 Such a MinMax series is a sequence of alternating minima and maxima (see Figure~\ref{fig:fdax}, where the minimal and maximal values
 are emphasized by a line).

\begin{figure} %1.4
  \centering
  \includegraphics[width=0.8\linewidth]{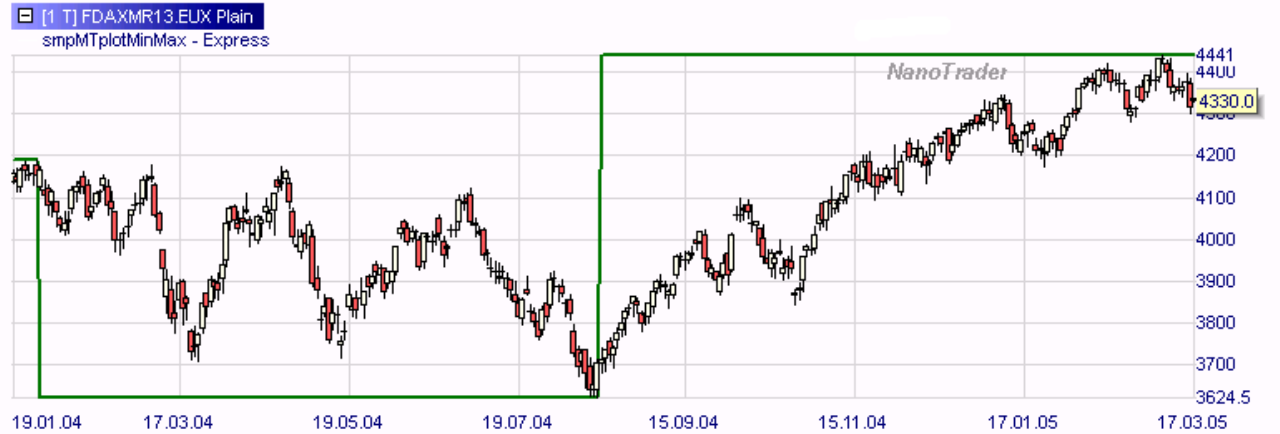}
  \caption{Chart of Figure~\ref{fig:fdax} with other MinMax series.}
  \label{fig:fdax_new_timescale}
\end{figure}

 The basic idea of the MinMax algorithm in \cite{Maier-Paape2013} to find a MinMax process are so called SAR (stop and reverse) processes which are indicators
 that can have only the two different values ``up'' or ``down''. For instance the MACD (moving average convergence/divergence) indicator of~\cite{Appel2005}
 can be used as SAR process. Simplified speaking, this process points up when the MACD series lies above its signal line and points down when its vice versa.
 Once the SAR process points up, the MinMax algorithm looks for a relevant maximum, and a relevant minimum is searched for, if the SAR process points down.
 See \cite{Maier-Paape2013} for details. Together with Definition~\ref{Trend_def}, trends can then be recognized automatically.

 Of course the MACD indicator is controlled by parameters (default are $12$, $26$ and $9$) which in turn affect the SAR process and therefore also the
 extrema-induced trends. To obtain a one-dimensional controlling parameter, a common factor called ``timescale'' is used in the MinMax algorithm
 of~\cite{Maier-Paape2013} that scales the above default parameters of the MACD indicator simultaneously in order to accelerate or slow down the accompanying
 SAR process. In the following sections we will use only the MACD to induce the SAR process and use only the timescale parameter to adjust the trend
 behavior. For instance, if the timescale parameter is enlarged, fewer extreme are recognized by the MinMax algorithm (see Figure~\ref{fig:fdax_new_timescale}).

 There is one subtlety, however: In order to avoid that the SAR process oscillates quickly in case MACD and signal line have multiple crossing points nearby,
 we adjust the SAR process slightly (for a change of the SAR process, the distance of MACD and signal line needs to be above some minimal threshold of
 $\delta = 0.3 \cdot \mbox{ATR} (100)$; see \cite[Subsection 2.1]{Maier-Paape2013} for details).

 As mentioned above we want to study statistics of trends. Therefore we first set the timescale parameter in section~\ref{sec:crosscorrelation} in a reasonable
 way. For doing so the ``wavelength'' of a chart is calculated in subsection~\ref{subsec:wavelength} via cross-/autocorrelation.
 Then in subsection~\ref{subsec:calibration} with the help of the wavelength the MinMax algorithm is calibrated. The calculated trends are then analyzed
 in section~\ref{sec:statisticalPropertiesTrend}.

 Although we did intensive literature research, we were not able to find closely related work on automatic trend analysis which used a similar geometric trend
 definition via ``MinMax processes''. Clearly, many other authors tried to design trading systems based on trends with miscellaneous results, see e.g.
 \cite{FB1966,Leuthold1972,Sidney1961}. However, they do not use such a concrete definition of trends but define only some parameter dependent filter rules to
 trade up and down movements.

\textbf{Acknowledgment} This paper was funded as Seed Fund Project 2011/2012, RWTH Aachen.

%%%%%%%%%%%%%%%%%%%% 2.

\section{Significant Period Length and Cross-Correlations}\label{sec:crosscorrelation} % 2.

 In the following we calculate the wavelength of a chart in subsection~\ref{subsec:wavelength} via cross-/auto\-correlation. Therefore we first introduce the
 corresponding notion and explain the calculation process. Afterwards the wavelength of some charts are presented. If the wavelength is known one can start
 calibrating the MinMax algorithm, see subsection~\ref{subsec:calibration}. Therefore the interrelation between the principal parameter (timescale) of the
 MinMax algorithm and the wavelength is detected.

%%% 2.1

\subsection{Wavelength of a Chart}\label{subsec:wavelength} % 2.1

 Our first goal is to get a notion of a ``wavelength'' of a time series. Intuitively the wavelength should stand for some sort of natural fundamental oscillation
 of the price process of a chart. In the sequel we will try to adjust the MinMax algorithm such that the period length of the resulting extrema matches
 the wavelength. Here the period length is the average distance between two consecutive minimal or two maximal values in time. To determine the wavelength of a
 chart we want to use the idea of cross-correlations.

 Assume $\mathbf{X}= (X_t)_{t=0,...,N}\in\R^{N+1} $ and $\mathbf{Y}= (Y_t)_{t=0,...,N}\in\R^{N+1}$ are two time series obtained from realizations of two
 stochastic processes. The empirical correlation $\phi\in[-1,1]$ of these two processes is then defined by
 \begin{align*}
   \phi := \textnormal{Corr\,}(\mathbf{X},\mathbf{Y}) :=
   \frac{\langle \mathbf{X},\mathbf{Y}\rangle}{\sqrt{\langle\mathbf{X},\mathbf{X}\rangle}\sqrt{\langle\mathbf{Y},\mathbf{Y}\rangle}},
 \end{align*}
 where $\langle\cdot,\cdot\rangle$ is the euclidean inner product.

 Similarly we define cross-correlation of a time series.

\begin{definition} \label{def:crosscorrelation} %def 2.1
  (Cross-correlation/Autocorrelation)\\
  Let $\mathbf{X} = \big(X_t\big)_{t\,\in\,\N_0}$ be the time series obtained from a stochastic process.
  We then call the empirical correlation of $\mathbf{Z}_1 = \big(X_t\big)_{t\,=\,0,\ldots,N}$ and $\mathbf{Z}_2:=\mathbf{Z}_2^n :=
  (X_{t+n})_{t=0,...,N}$ for given $n,N\in\N$ the {\bfseries cross-} or {\bfseries autocorrelation} of $\mathbf{X}$ with time shift $n$.
\end{definition}

 In order to use the concept of the cross-correlation for the introduction of a wavelength of charts, we need to do some transformations.

 A chart typically is of the form of a candle or bar chart, i.e. for one instant of time (one candle/bar) we get four values (open, close, high and low).
 The values open and close are the market values at a specific time, which are more or less randomized. However the values high and low are the maximum
 and minimum of a small period of time and therefore represent more than just one market price at a predefined point in time. For this reason we will use
 $\nicefrac{(\text{high} + \text{low})}{2}$ to get one value for each candle/bar, i.e. we use
 \begin{align} \label{NR3-1} %NR_2.1
    a_t &:=\frac{\text{high}(t) + \text{low}(t)}{2},
    && t\in\N_0
    && \text{(used value for the }t\text{-th candle of the chart),}
 \end{align}
 as real valued time series representing the price process of the chart. In case \eqref{NR3-1} has some sort of a dominant wavelength $n^*$ after shifting
 $a_t$ by $n^*$ periods to the right (at least on average) maxima should more or less be close to maxima of the unshifted series and minima should be close
 to minima.

 Unfortunately, even if this happens, we cannot directly measure that with the cross-correlation, since for instance two overlayed (positives) minima give a
 small contribution in the cross-correlation of Definition~\ref{def:crosscorrelation}, whereas two overlayed maxima give a large contribution.

 Therefore we have to do some modifications to obtain a large cross-correlation for the case that two maxima and minima clash and a small cross-correlation,
 if maxima hit minima. In order to obtain this behaviour, we subtract from \eqref{NR3-1} a moving average, such that an alleged minimum will be negative and
 an alleged maximum will be positive for the resulting series.

\begin{figure} %fig_2.1
   \centering
   \includegraphics[width=0.6\linewidth]{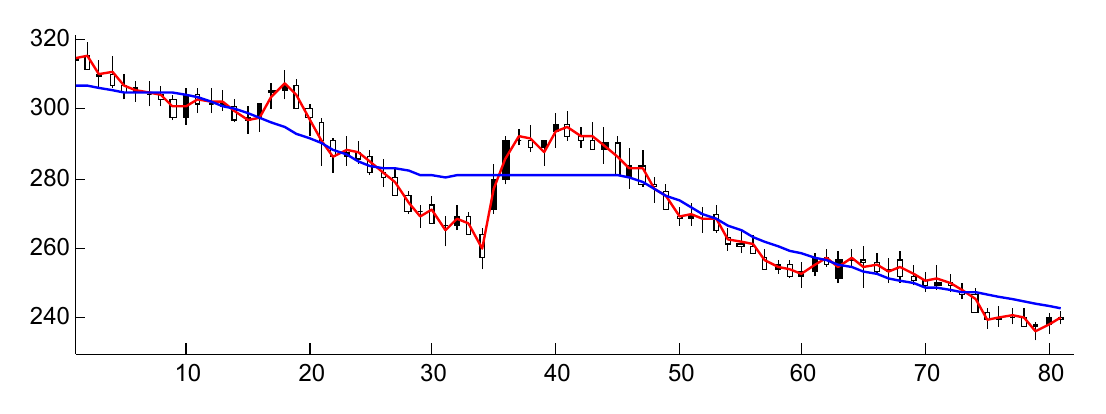}
   \caption{A chart with its values $a_t$ (red line) and $b_t^n$, $n=20$ (blue line).}
   \label{fig:chart}
\end{figure}

\begin{figure} %fig_2.2
   \centering
   \includegraphics[width=0.6\linewidth]{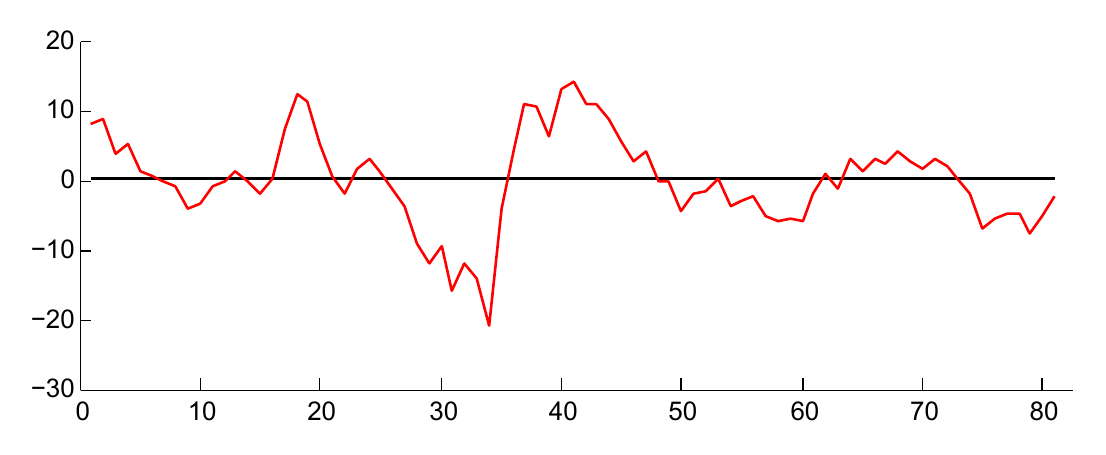}
   \caption{Plot of $X_t^n=a_t-b_t^n$, $n=20$, with $a_t$ and $b_t^n$ from Figure~\ref{fig:chart}.}
   \label{fig:chart_diff}
\end{figure}

 Of course the resulting maxima and minima of the new series will depend on the amount of periods we use for the moving average. It would be meaningful to take
 the moving average of the chart with a span of $n$ candles, where $n$ is equal or close to the dominant wavelength. However, since we do not know this dominant
 wavelength yet, we use $n$ for now as a parameter. To compute the average at one candle we will therefore take $\nicefrac{n}{2}$ periods in front of and
 $\nicefrac{n}{2}$ after the current candle. The difference of the real valued time series $a_t$ and its moving average becomes our time series $\mathbf{X}$,
 i.e. for
 \begin{align*}
   b_t^n &:=\left(\frac{1}{2\lfloor\nicefrac{n}{2}\rfloor +1}\sum_{i=-\lfloor \nicefrac{n}{2}\rfloor}^{\lfloor\nicefrac{n}{2}\rfloor} a_{t+i}\right)
        &&\text{(averaging of $a_t$ for $t\geq\lfloor\nicefrac{n}{2}\rfloor$)}
 \end{align*}
 we set
 \begin{align}\label{eq:X_t} %NR_2.2
   X_t^n &:=a_t - b_t^n.
 \end{align}
 If our time series $(a_t)_{t=0,...,M-1}$ is of length $M$, we now can compute the series
 $\mathbf{X}^n:=(X^n_t)_{t=\lfloor\nicefrac{n}{2}\rfloor,...,M-1-\lfloor\nicefrac{n}{2}\rfloor}$ of length $M-2\lfloor\nicefrac{n}{2}\rfloor$.
 In Figure~\ref{fig:chart} one can see a chart, where the red line are the $a_t$ and the blue line $b_t^n$ for $t=1,...,81$ and $n=20$, where for the
 purpose of illustration our series $a_t$ is defined for $t\in\Z$. Since all values are positive, we obtain $X_t^n$ oscillating around zero, see
 Figure~\ref{fig:chart_diff}.

 Now the cross-correlation will be large if we overlay two maxima or two minima (which now can be negative) in Definition~\ref{def:crosscorrelation}.

\begin{definition}
   (``Cross-correlation'' of a time series)\\
   Let $A=(a_t)_{t=0,...,M-1}$ be a real valued time series and $n\in\N$ be fixed. We define the cross-correlation of $A$ with time shift $n$ as
   \begin{align*}
     \phi_n := \textnormal{Corr\,}(\mathbf{Z}_1^n,\mathbf{Z}_2^n),
   \end{align*}
   where $\textbf{Z}_1^n := (X^n_{t})_{t=\lfloor\nicefrac{n}{2}\rfloor,...,N-\lfloor\nicefrac{n}{2}\rfloor}$ and $\textbf{Z}_2^n :=
   (X^n_{t+n})_{t=\lfloor\nicefrac{n}{2}\rfloor,...,N-\lfloor\nicefrac{n}{2}\rfloor}$ and $X_t^n$ as in~\eqref{eq:X_t}.
   For a given shift $n$ we choose $N$ maximal, i.e. we use $N:=M-n-1$.
\end{definition}

 The $n$ with the largest cross-correlation will be called the \textbf{dominant wavelength} $n^*$ of the signal. In the next subsection
 the MinMax algorithm is adjusted such that it reproduces this dominant wavelength $n^*$.

\begin{figure}[p] %fig_2.3
  \centering
  \includegraphics[width=0.99\linewidth]{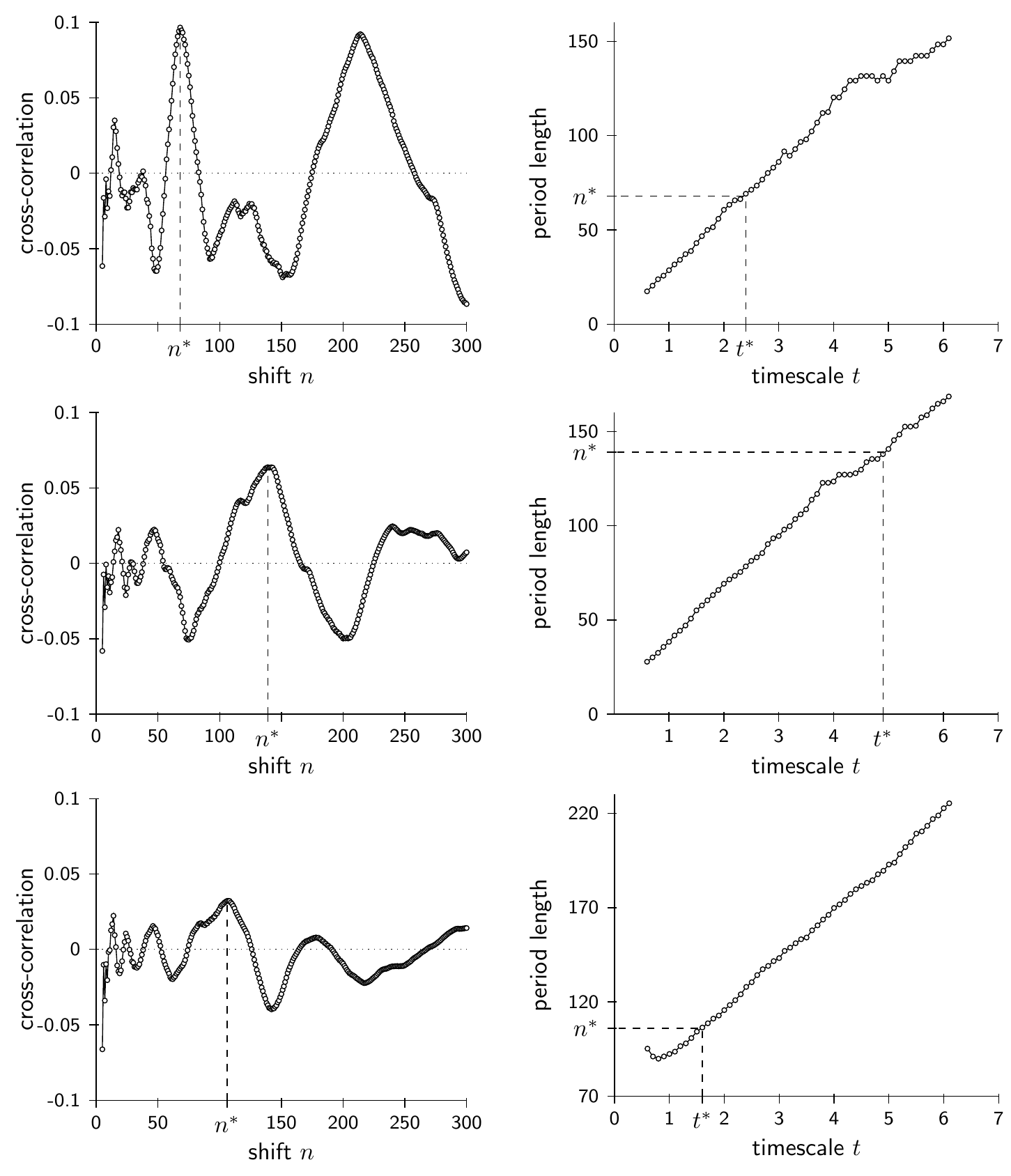}
  \caption{Euro US-Dollar: In the left column one sees the cross-correlation for values of $n$ between $1$ and $300$ and in the right column we
  plot the average period length one gets from the MinMax algorithm of \cite{Maier-Paape2013} for different values of the timescale parameter.
  The aggregation is 1 day, 1 hour and 10 minutes from top to bottom.  $n^*$ is the dominant wavelength and $t^*$ the corresponding timescale
  from Table~\ref{tab:wavelength} and \ref{tab:timescale_from_wavelength}, respectively.}
  \label{fig:eur_usd}
\end{figure}

 In the left column of Figure~\ref{fig:eur_usd} one sees the cross-correlation of the Chart of EUR-USD with different aggregations. The evaluation period used
 for the following studies is given in Table~\ref{tab:period_of_time}. From Figure~\ref{fig:eur_usd} one approximately can extract the dominant wavelength, e.g.
 in the figure belonging to the day Chart of EUR-USD, the biggest extrema at $n^*=68$ probably marks the wavelength. In fact for $n= 68$ the cross-correlation
 is largest with value $0.0966$. The range for lower $n$ shows oscillations, which are not meaningful. Similarly, the last big extremum which shows up between
 $n=200$ and $n=250$ corresponds to a multiple of the dominant wavelength of $n^*=68$. We proceed analogously with the remaining time units of EUR-USD in
 Figure~\ref{fig:eur_usd}. The corresponding dominant wavelengths were collected in Table~\ref{tab:wavelength}. In addition in Table~\ref{tab:wavelength} are
 wavelengths for some other underlyings as well. Please note that although the cross-correlations are of small size there are still strong oscillations.
 Therefore one can assume that these values are meaningful.

\begin{table}[ht] %tab_2.1
  \centering
  \caption{Some dominant wavelengths $n^*$ determined by the cross-correlation (evaluation time is given by Table~\ref{tab:period_of_time}).}
  {\footnotesize %
    \newcommand{\mc}[3]{\multicolumn{#1}{#2}{#3}}
    \begin{tabular}{rrrrrrrrrrrrr}
      \toprule
        && \mc{2}{c}{EUR-USD} && \mc{2}{c}{FDAX} && \mc{2}{c}{Gold} && \mc{2}{c}{Crude Oil}\\
           \cmidrule{3-4}  \cmidrule{6-7} \cmidrule{9-10}  \cmidrule{12-13}
      Aggregation  && $n^*$ & Corr && $n^*$ & Corr && $n^*$ & Corr && $n^*$ & Corr\\
      \midrule
      1d           &&  68 & 0.0966 && 101 & 0.0415 &&  59 & 0.2550 &&  86 & 0.0160\\
      1h           && 139 & 0.0637 && 157 & 0.0600 && 112 & 0.0407 &&  97 & 0.0401\\
      10min        && 106 & 0.0322 &&  59 & 0.0799 && 111 & 0.0701 &&  64 & 0.0026\\
      \bottomrule
    \end{tabular}
  }%
  \label{tab:wavelength}
\end{table}

\begin{table}[ht] %tab_2.2
  \centering
  \caption{Starting date for all studies (terminal date is always 25.01.2013). All historical data are from FIDES.}
  {\footnotesize %
    \newcommand{\mc}[3]{\multicolumn{#1}{#2}{#3}}
    \begin{tabular}{rcccc}
      \toprule
        & EUR-USD & FDAX & Gold & Crude Oil\\
      \midrule
      1d           & 15.07.1985 & 02.08.1993 & 14.09.1990 & 14.09.1990\\
      1h           & 14.07.2009 & 17.12.1999 & 11.07.2005 & 29.11.2004\\
      10min        & 03.01.2011 & 03.01.2011 & 03.01.2011 & 03.01.2011\\
      \bottomrule
    \end{tabular}
  }%
  \label{tab:period_of_time}
\end{table}

%%%

\subsection{Calibration of the MinMax algorithm}\label{subsec:calibration} % 2.2

 To adjust the trend finder according to the dominant wavelength $n^*$ given by subsection~\ref{subsec:wavelength} it is necessary to detect the
 interrelation between the principle parameter (timescale) of the MinMax algorithm and the wavelength. One easy approach is to vary this
 parameter and compute for each fixed adjustment the averaged period length given by the relevant minima and maxima which we get from the trend
 finder. We will vary the timescale between $0.4$ and $6.0$ in $0.1$ steps. If we plot the resulting period length against the timescale, we get
 the result shown in the right column of Figure~\ref{fig:eur_usd} for EUR-USD. In all three aggregations the dependence between timescale and
 observed average period length is close to (affine) linear. Now we can read the parameter for a given wavelength we got from
 subsection~\ref{subsec:wavelength}, see left column of Figure~\ref{fig:eur_usd} for EUR-USD and Table~\ref{tab:wavelength}.
 The values found for the timescale parameter $t^*$ that matches $n^*$ best can be found in Table~\ref{tab:timescale_from_wavelength}.

\begin{table}[ht] %tab_2.3
  \centering
  \caption{Choices for the timescale parameter $t^*$ to meet the dominant wavelength $n^*$.}
  {\footnotesize %
    \begin{tabular}{rcccc}
      \toprule
      Aggregation  & EUR-USD & FDAX & Gold & Crude Oil\\
      \midrule
      1d           & 2.4 & 3.7 & 2.2 & 3.4\\
      1h           & 4.9 & 6.0 & 3.8 & 3.8\\
      10min        & 1.6 & 1.6 & 2.5 & 1.6\\
      \bottomrule
    \end{tabular}
  }%
  \label{tab:timescale_from_wavelength}
\end{table}

 The idea of calibrating the MinMax algorithm with the dominant wavelength of the chart fixes the relevant extrema and therefore also the ``relevant'' trends.
 We call this procedure according to \cite{Maier-Paape2013} the 1--2--3 trend indicator. In the next section we want to verify the quality of this setting from
 a practical point of view. For example we can verify how long trends are active, i.e. how many maxima and minima are contained in the trends. The more maxima
 and minima we have for a trend on average the better the 1--2--3 trend indicator seems to work, because a trend with only two maxima and two minima is not
 helpful, because it disappears as quickly as it appeared. Before analysing basic statistical properties of trends we should do some remarks.

\begin{remark} %rem_2
  \begin{enumerate}
    \item The other local maxima in the plots of the cross-correlation of Figure~\ref{fig:eur_usd} correspond to other -- not
          so significant -- wavelengths of the chart. After determining the respective timescales several in one other nested
          trends can be calculated. The study of these nested trends will however not be the subject of this paper.
    \item As alternative way to determine relevant wavelengths/frequencies of a chart the empirical mode decomposition (EMD)~\cite{HSL+1998}
          could be used. D\"urschner used in~\cite{Duerschner2013} the EMD to analyse financial time series. Nevertheless, this method would
          not select a dominant wavelength.
    \item Fourier Analysis and its (fast) Fourier transformation also provides a ``frequency decomposition'' and with it ``strong modes'',
          see e.g.~\cite{MV2006}. However, the comparison with this approach would be beyond the scope of this paper.
    \item In this section we determined the cross-correlation from all available data. Alternatively one could calculate the cross correlations
          at each time $t$ from the last $T$ series points $(a_s)_{s= t-T+1,...,t}$. This way the cross-correlation plots as well as $n^*$ and
          $t^*$ would depend on the time $t$, which might be more adapted to the actual market phase.
  \end{enumerate}
\end{remark}

%%%%%%%%%%%%%%%%%%%%

\section{Basic Statistical Properties of Trends}\label{sec:statisticalPropertiesTrend} % 3.

 Since we already fixed the parameter of the 1--2--3 trend indicator for each underlying, we now want to study trends in detail.
 The question is what kind of information about trends are important. Before we try to generate informations from trends which can help us for
 investment decisions we first should evaluate the quality of the detected trends. Therefore we define in subsection~\ref{subsec:trend_quality}
 two important measurement numbers, the dynamic and the lifetime of a trend. First we compare the corresponding results for different settings of
 the timescale on each underlying. If this analysis shows that the detected trends are meaningful we have a proper statistical basis of trends
 to work with. Therefore we analyze on this statistical basis complex chance and risk measurements in subsection~\ref{subsec:add_properties} and
 discuss the expected values of these measurement numbers for each underlying using the fixed timescale from Table~\ref{tab:timescale_from_wavelength}.
 In subsection~\ref{subsec:pos_p2} we then focus on the distribution of the dynamic to get a more detailed view on the shape of a trend.

%%%

\subsection{How to measure trend quality}\label{subsec:trend_quality} % 3.1

 There are different methods to measure the strength of a trend. Classical methods to measure trend strength are e.g. the random walk
 index~\cite{Poulos1991}, the Aroon indicator~\cite{CK1994} and the relative strength index (RSI)~\cite{Wilder1978,Heckmann2009}.
 All these indicators are applied directly to the price data of the chart and do not use our trend definition with P1, P2 and P3 explicitly.
 In this paper we therefore use the \textbf{dynamic} and the \textbf{life-time} of a trend.

\begin{figure}[b] %fig_3.1
  \centering
  \includegraphics[width=0.5\linewidth]{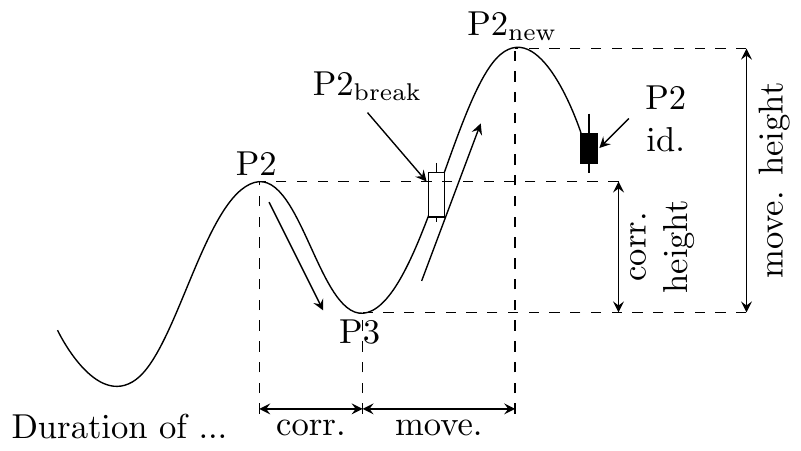}
  \caption{Situation of type 2--3--2 of Definition~\ref{def:trend_properties}.}
  \label{fig:trend2}
\end{figure}

\begin{definition}\label{def:dynamic} %def_3.1
  (Dynamic of a trend)\\
  By \textbf{dynamic} we denote
  \begin{align*}
    \textnormal{dynamic} := \frac{\textnormal{move. height}}{\textnormal{corr. height}}
                          = \frac{|\textnormal{P2}_{\textnormal{new}}-\textnormal{P3}|}{|\textnormal{P2}-\textnormal{P3}|} > 1,
  \end{align*}
  i.e. the quotient of the height of the movement and the height of the preceding correction phase, see Figure~\ref{fig:trend_definition}
  or Figure~\ref{fig:trend2}. Later for the statistics we always use the empirical expectation $\mathbb{E}(\textnormal{dynamic})$.
\end{definition}

\begin{remark}\label{remark:lag} %rem_3.
  In Figure~\ref{fig:trend2} the last candle is marked with $\textnormal{P2 id.}$, which means that $\textnormal{P2}_{\textnormal{new}}$ is identified.
   Extrema are identified when the ``status'' series of the MinMax algorithm  has changed, which of course is after the extrema occurred
   (see \cite{Maier-Paape2013} for details).
\end{remark}

 The dynamic is very meaningful because it sets the movement phase in context to the correction phase. If this quantity takes a high value it is a good
 signal for the trend quality because this means that during the movement phase the price was increasing much and during the correction phase the price
 was decreasing less. In fact this is what a trader wants.

\begin{figure}[p] %fig_3.2
  \centering
  \includegraphics[width=0.99\linewidth]{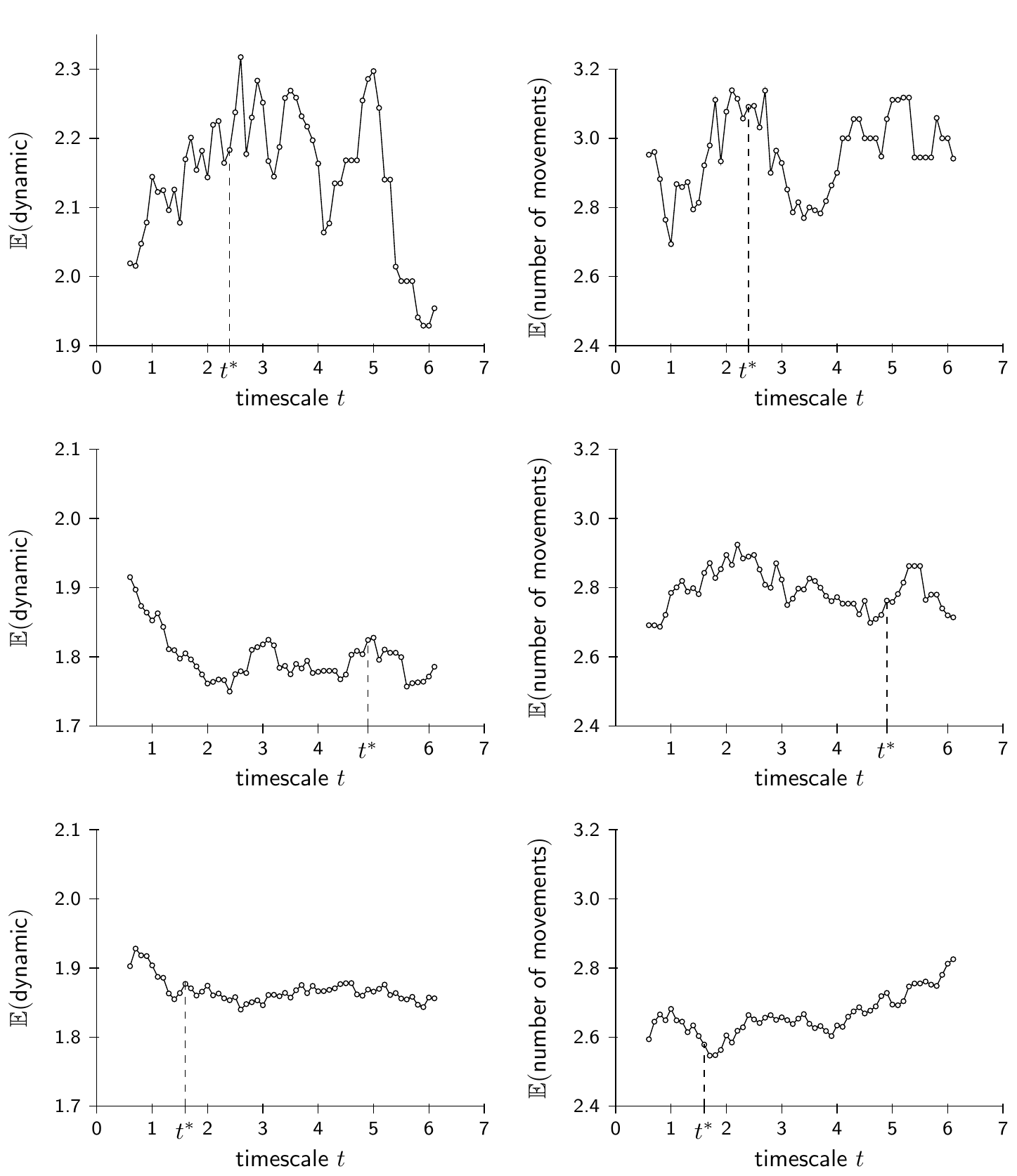}
  \caption{Euro US-Dollar: In the left column one sees the timescale plotted against the dynamic and in the right one sees
  the timescale against the number of movements. The aggregation is 1 day, 1 hour and 10 minutes from top to bottom and $t^*$
  is the timescale from Table~\ref{tab:timescale_from_wavelength}.}
  \label{fig:eur_usd_dynamic_movements}
\end{figure}

 In the left column in Figure~\ref{fig:eur_usd_dynamic_movements} the dynamic against the parameter timescale is shown. For day aggregation we see that
 the dynamic takes its maximal value of $2.31$ at a timescale of $2.6$. This timescale is nearly identical with $t^*$ we calculated from the dominant
 wavelength in section~\ref{subsec:wavelength}. For the other aggregations the values do not match that good, but if one ignores the values for timescale
 below one, the dynamic does not vary that much.

 Another useful criterion for evaluating trends could be the \textbf{life-time of a trend}. This does not have a clear definition, e.g. it could mean the
  time under which the trend exists. This view seems not so meaningful because trends on a higher time scale would then usually live longer then those on
   a lower time scale. However, we will focus on characteristics of trends, that can be compared amongst various time scales. We therefore use the following
   definition.

\begin{definition} %def_3.2
  (Life-time of a trend)\\
  We say a trend is long living if it consists of a lot of extremal values. We quantify this by the number of movements.
  This number is always at least two, since we also count the initial movement from \textnormal{P1} to \textnormal{P2},
  see Figure~\ref{fig:trend_definition}.
\end{definition}

 The reason for this point of view is that of course trends with a high number of movements are most meaningful for trading strategies.
 Thus we plot in the right column of Figure~\ref{fig:eur_usd_dynamic_movements} the average number of movements per trend phase against the timescale.
 For the aggregation of 1 day and 1 hour the values $t^*$ for the timescale in Table~\ref{tab:timescale_from_wavelength} yields a good number of movements.
 For 10 minutes the number of movements does not vary much anyway.

 So far it seems that our trend finder with the settings calculated from the wavelength is meaningful and therefore we will continue in the sequel always
 with the trend finder with timescale $t^*$.

%%%

\subsection{Additional trend properties}\label{subsec:add_properties} % 3.2

 We are now interested in collecting some basic properties of our trends which can help investors for investment decisions.
 For this reason we have to make some definitions.

\begin{definition}\label{def:trend_properties} %def_3.3
  (Some properties of a trend)\\
  The average true range (\textnormal{ATR}), see~\cite{Wilder1978}, is the moving average of the true range
  \begin{align*}
    \textnormal{TR}:=\max\{\textnormal{high} - \textnormal{low},\textnormal{high}-\textnormal{close}[1],\textnormal{close}[1]-\textnormal{low}\},
  \end{align*}
  where high and low means the high and low of the current period, respectively and close$[1]$ means the close of the previous period.
  Here the average of $100$ periods has been used. Assuming that a large average true range, i.e. a high volatility, leads to large movements and
  vice versa we use this value for normalization.

  In the following we will distinguish between three different situations.
  \begin{description}
    \item[1--2--3:]
      We will call situations with identified points \textnormal{P1}, \textnormal{P2} and \textnormal{P3} an 1--2--3, see Figure~\ref{fig:trend_situations}
      left. In these situations no trend is active yet. If \textnormal{P3} is identified only after or during the break of the last \textnormal{P2},
      i.e. there is an active trend, we will ignore the situation, because such situations are not of use-oriented interest relating to the breakout
      of \textnormal{P2}.

      The number $\#(\textnormal{1--2--3})$ counts the number of situations which occur in the chart. We can use this property e.g. for forecasting how
      likely a trend will be generated after the occurrence of an 1--2--3.

      Next we define
      \begin{align*}
        R_{\textnormal{1--2--3},\textnormal{ATR}}
          &:= \frac{R_{\textnormal{1--2--3}}}{\textnormal{ATR(P3)}} := \frac{\textnormal{close}(\textnormal{P3 ident.})-\textnormal{P3}}{\textnormal{ATR(P3)}}
          &&\text{(risk)},\\
        G_{\textnormal{1--2--3},\textnormal{ATR}}
          &:= \frac{G_{\textnormal{1--2--3}}}{\textnormal{ATR(P3)}} := \frac{\textnormal{P2}-\textnormal{close}(\textnormal{P3 ident.})}{\textnormal{ATR(P3)}}
          &&\text{(goal)},\\
        R_{\textnormal{1--2--3},\%}
          &:= \frac{R_{\textnormal{1--2--3}}}{R_{\textnormal{1--2--3}}+G_{\textnormal{1--2--3}}} := \frac{\textnormal{close}(\textnormal{P3 ident.})-
              \textnormal{P3}}{\textnormal{P2}-\textnormal{P3}}
          &&\text{(risk in \%)},\\
        G_{\textnormal{1--2--3},\%}
          &:= \frac{G_{\textnormal{1--2--3}}}{R_{\textnormal{1--2--3}}+G_{\textnormal{1--2--3}}} = 1 - R_{\textnormal{1--2--3},\%}
          &&\text{(goal in \%)},
      \end{align*}
      which have to be positive for long-situations, see Figure~\ref{fig:trend_situations} left. For short-situations, we use the absolute values of these
      expressions. The idea is that the occurrence of an 1--2--3 can be a trigger for a position entry suspecting a new trend with \textnormal{P1} as stop
      loss level and the last \textnormal{P2} as the smallest meaningful target.

\begin{figure} %fig_3.3
  \centering
  \includegraphics[width=0.8\linewidth]{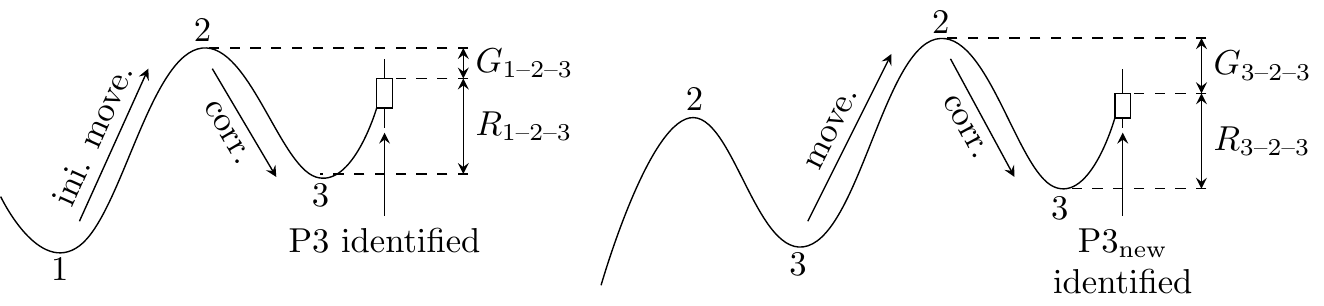}
  \caption{Situations of type 1--2--3 (left) and 3--2--3 of Definition~\ref{def:trend_properties}.}
  \label{fig:trend_situations}
\end{figure}

    \item[3--2--3:]
      Analogously we will call situations with identified points \textnormal{P3}, \textnormal{P2} and $\textnormal{P3}_{\textnormal{new}}$ a 3--2--3,
      see Figure~\ref{fig:trend_situations} right. In these situations a trend is already active, but the last \textnormal{P2} is not broken when the
      new \textnormal{P3} has been fixed. Otherwise we again ignore the situation for the same reason as for the 1--2--3 case.

      The number of such situations are denoted by $\#(\textnormal{3--2--3})$ and we set again
      \begin{align*}
        R_{\textnormal{3--2--3},\textnormal{ATR}}
          &:= \frac{R_{\textnormal{3--2--3}}}
                   {\textnormal{ATR}(\textnormal{P3}_{\textnormal{new}})}
           := \frac{\textnormal{close}(\textnormal{P3}_{\textnormal{new}}\textnormal{ ident.})-\textnormal{P3}_{\textnormal{new}}}
                   {\textnormal{ATR}(\textnormal{P3}_{\textnormal{new}})}
           &&\textnormal(risk),\\
        G_{\textnormal{3--2--3},\textnormal{ATR}}
          &:= \frac{G_{\textnormal{3--2--3}}}
                   {\textnormal{ATR}(\textnormal{P3}_{\textnormal{new}})}
           := \frac{\textnormal{P2}-\textnormal{close}(\textnormal{P3}_{\textnormal{new}}\textnormal{ ident.})}
                   {\textnormal{ATR}(\textnormal{P3}_{\textnormal{new}})}
           &&\textnormal(goal),\\
        R_{\textnormal{3--2--3},\%}
          &:= \frac{R_{\textnormal{3--2--3}}}{R_{\textnormal{3--2--3}}+G_{\textnormal{3--2--3}}} :=
              \frac{\textnormal{close}(\textnormal{P3}_{\textnormal{new}}\textnormal{ ident.}) -
              \textnormal{P3}_{\textnormal{new}}}{\textnormal{P2}-\textnormal{P3}_{\textnormal{new}}}
          &&\text{(risk in \%)},\\
        G_{\textnormal{3--2--3},\%}
          &:= \frac{G_{\textnormal{3--2--3}}}{R_{\textnormal{3--2--3}}+G_{\textnormal{3--2--3}}} = 1 - R_{\textnormal{3--2--3},\%}
          &&\text{(goal in \%)},
      \end{align*}
      which are positive for long-situations and otherwise we use the absolute values, see Figure~\ref{fig:trend_situations} right.

    \item[2--3--2:]
      Situations where we know the points \textnormal{P2}, \textnormal{P3} and the consecutive $\textnormal{P2}_{\textnormal{new}}>\textnormal{P2}$
      of an active trend are called 2--3--2, see Figure~\ref{fig:trend2}. In these situations, we do not know the next minimum value, which might be
      a new \textnormal{P3} or -- if below the old \textnormal{P3} -- leads to a break of the trend.

      In Definition~\ref{def:dynamic} we already defined the dynamic of a trend. Another question is how much time is needed for a movement phase
      with respect to the time for the preceding correction phase. For this reason we first define the points with an additional time parameter.
      We say that \textnormal{P2} occurred at time $t2$, \textnormal{P3} at time $t3$ and $\textnormal{P2}_{\textnormal{new}}$ at time
      $t2_{\textnormal{new}}$. We then define
      \begin{align*}
        \textnormal{relative duration of dynamic}
        := \frac{t2_{\textnormal{new}}-t3}{t3-t2}.
      \end{align*}
      Since $\textnormal{P2}_{\textnormal{new}}$ is always identified with a time lag, see Remark~\ref{remark:lag}, we are also interested in the effects
      of this delay. Therefore we define the lagged dynamics similar to the dynamic but we replace the point
      $(t2_{\textnormal{new}},\textnormal{P2}_{\textnormal{new}})$ with $(t2_{\textnormal{new,id}},
            \textnormal{close}(\textnormal{P2}_{\textnormal{new}}\textnormal{ ident.}))$, where $t2_{\textnormal{new,id}}$ is the time when
      $\textnormal{P2}_{\textnormal{new}}$ is identified, i.e. we have
      \begin{align*}
        \textnormal{lag. dyn}
          &:= \frac{|\textnormal{close}(\textnormal{P2}_{\textnormal{new}}\textnormal{ ident.})-\textnormal{P3}|}{|\textnormal{P2}-\textnormal{P3}|}\\
                     \textnormal{rel. duration of lag. dyn}
          &:= \frac{t2_{\textnormal{new,id}}-t3}{t3-t2}.
      \end{align*}
      We note that the lagged dynamic does not have to be larger than one. These quantities are important because the identification point
      $t2_{\textnormal{new,id}}$ is immediately known as it occurs. In contrast the point $t2_{\textnormal{new}}$ always lies in the past
      when $\textnormal{P2}_{\textnormal{new}}$ is recognized and can therefore not be used directly.

      Another important point is the break of $\textnormal{P2}$. We denote the first candle which breaks $\textnormal{P2}_{\textnormal{break}}$, i.e.
      whose highest value is larger than $\textnormal{P2}$, by the point $(\textnormal{t2}_{\textnormal{break}},\textnormal{P2}_{\textnormal{break}})$,
      see Figure~\ref{fig:trend2}. As above we define
      \begin{align*}
        \textnormal{rel. duration of break}
          &:= \frac{t2_{\textnormal{break}}-t3}{t3-t2}.
      \end{align*}
  \end{description}
\end{definition}

\begin{remark} %rem_4.
  \begin{enumerate}
    \item We note that in both situations 1--2--3 as well as 3--2--3 there are possible countertrends when the succeeding situation
          breaks the last \textnormal{P3}.
    \item The probability that a trend breaks after a 2--3--2 can be computed by $\nicefrac{\textnormal{\#trends}}{\textnormal{\#(2--3--2)}}$,
          because the number of observed breaks equals the number of trends $\textnormal{\#trends}$ (each trend breaks exactly one time) and
          $\textnormal{\#(2--3--2)}$ gives us the number of all observed situations. Thus the probability of pass $\textnormal{P2}_{\textnormal{new}}$
          after a 2--3--2 can be computed by
    \begin{align*}
      1 - \frac{\textnormal{\#trends}}{\textnormal{\#(2--3--2)}}.
    \end{align*}
  \end{enumerate}
\end{remark}

 Several properties of a trend are calculated for EUR-USD, DAX-Future, Gold and Crude Oil on different time-units, see Table~\ref{tab:other_critera_eurusd_fdax}
 and Table~\ref{tab:other_critera_gold_oil}. We tried to calculate them over a possibly huge period of time. Let us again mention that we used the
 dominant wavelength calculated in section~\ref{subsec:wavelength} for calibrating the MinMax process and hence the 1--2--3 trend indicator, i.e. we used
 the timescale $t^*$ of Table~\ref{tab:timescale_from_wavelength}. Now we discuss the results for the quantities of Definition~\ref{def:trend_properties}
 starting with the 1--2--3 situations.

\paragraph{1--2--3:}
 From Tables~\ref{tab:other_critera_eurusd_fdax} and~\ref{tab:other_critera_gold_oil} we extract that for the highest time-unit, day base, only 44 situations
 of type 1--2--3 according to Definition~\ref{def:trend_properties} on the EUR-USD where found and only 16, 44 and 25 situations on the DAX-Future, Gold and
 Crude Oil respectively. In contrast there were 206 situations for the smallest time-unit on the EUR-USD and 206, 198 and 453 situations on the other underlyings
 within the evaluation period of our data. Of course the reason is the number of periods/candles used for the evaluations, which is lowest on day basis.
 Therefore we find most situations in the lowest time-unit.

 The probability of activating a trend after a 1--2--3 situation is identified varies between only 28\% and 62\% on day basis and between 42\% and 55\% for
 intraday data. Of course the reason for the fluctuations of this value on day basis could be the small number of situations. Roughly speaking we have a 50\%
 chance that a trend will occur after a 1--2--3.

 The expectation of the risk $R_{\textnormal{1--2--3},\textnormal{ATR}}$ is always slightly higher or very close to the gain
 $G_{\textnormal{1--2--3},\textnormal{ATR}}$. This is confirmed by the relative risk $R_{\textnormal{1--2--3},\%}$, which is between 50\% and 60\%.

 The last quantity for 1--2--3 situations is the expectation of the ratio of the height of the correlation and the height of the initial movement.
 This value is about 70\%, which means that 70\% of the initial price movement gets lost during the correction phase.

\paragraph{3--2--3:}
 The number of 3--2--3 situation is in all underlyings always observed smaller than the number of 1--2--3 situations. The probability of passing P2
 after a 3--2--3 fluctuates between 28\% and 65\%, see Tables~\ref{tab:other_critera_eurusd_fdax} and~\ref{tab:other_critera_gold_oil}.

 The values for the risk $R_{\textnormal{3--2--3},\textnormal{ATR}}$ and $R_{\textnormal{3--2--3},\%}$ and the gain $G_{\textnormal{3--2--3},\textnormal{ATR}}$
 and also the quotient of the height of the correlation and the height of the preceding movement are very similar to the 1--2--3 setting.

\paragraph{2--3--2:}
 The probability of reaching a new P2 exceeding the old P2 under the condition that only the old P2, which is $\textnormal{P2}_{\textnormal{new}}$ in
 Figure~\ref{fig:trend2}, is identified by the 1--2--3 trend indicator increases significantly with rising the time-unit for the EUR-USD and also Crude
 Oil. This effect does not show up for the DAX-Future and Gold. For the Gold this quantity varies marginally with the time-units, so that the changes
 could be statistical fluctuations.

 The dynamic gives us the quotient of the height of the movement and correction phase. A high dynamic means, that the market price will cover a large
 distance during the movement phase compared to the correction phase, which is more favorable. We see that the average dynamic takes its highest value
 for all underlyings except Crude Oil on day basis with a value above 2 in these three cases. In any case the average dynamic is above 1.75 which is
 also reasonably large.

 For the relative duration of the dynamic, which is the quotient of the time needed for the movement phase and the time for the correction phase, this
 behavior is similar for the EUR-USD and the DAX-Future. The relative duration of the dynamic is much higher than the dynamic itself on all four
 underlyings, which means that the absolute value of the slope of the price changes is significantly higher for the correction than for the movement.
 This result is remarkable since movements are generally expected to be fast compared with corrections.

 From the two quantities for the lagged dynamic one sees that the time between $\textnormal{P2}_{\textnormal{new}}$ and the identification of this point,
 which in general is part of the new correction, is larger than the time for the preceding correction phase. The lagged dynamic itself is about one,
 which means that the price of the chart on average goes back to the level of the old P2 before $\textnormal{P2}_{\textnormal{new}}$ is identified. Thus
 even the price changes after $\textnormal{P2}_{\textnormal{new}}$ are slightly slower than during the preceding correction.

 The next two quantities $\mathbb{E} (\nicefrac{\textnormal{move. height}}{\textnormal{ATR}(\textnormal{P}3)})$ and
 $\mathbb{E} (\nicefrac{\textnormal{corr. height}}{\textnormal{ATR}(\textnormal{P}3)})$ in Tables~\ref{tab:other_critera_eurusd_fdax}
 and~\ref{tab:other_critera_gold_oil} set the height of the movement, respectively the correction in context to the average true range,
 see Definition~\ref{def:trend_properties}, at the last P3. The higher the value of
 $\mathbb{E} (\nicefrac{\textnormal{move. height}}{\textnormal{ATR}(\textnormal{P}3)})$ and the lower
 $\mathbb{E} (\nicefrac{\textnormal{corr. height}}{\textnormal{ATR}(\textnormal{P}3)})$ the better it is, because this means that the prices increased
 significantly during the movement phases and does not decrease much during the correction phase. We see that on average the height of the correction
 phase is slightly below 6 times and the height of the movement phase about 10 times the average true range. A very similar relation between these two
 values can be seen for
 $\mathbb{E} (\nicefrac{\textnormal{move. height}}{\textnormal{low}(\textnormal{P}3)})$ and
 $\mathbb{E} (\nicefrac{\textnormal{corr. height}}{\textnormal{low}(\textnormal{P}3)})$.
 One also can conclude that the DAX-Future and Crude Oil are comparably volatile, but Gold is less volatile and EUR-USD is the least volatile underlying.

 Interestingly, on every time-unit trends in all underlyings have only very few movements numbers, which is slightly below $3$ on average.

\begin{table}[p] %tab_3.1
  \centering
  \caption{Some quality criteria for EUR-USD and DAX-Future with different aggregations.}
  {\footnotesize
    \newcommand{\mc}[3]{\multicolumn{#1}{#2}{#3}}
    \newcommand{\mr}[3]{\multirow{#1}{#2}{#3}}
    \newcommand{\mrN}[1]{\multirow{2}{*}{#1}}
    \newcommand{\mrA}[1]{\multirow{1}{*}{#1}}
    \begin{tabular}{lcccccc}
      \toprule
      underlying &\mc{3}{c}{EUR-USD} &\mc{3}{c}{DAX-Future}\\
      \cmidrule(lr){2-4} \cmidrule(l){5-7}
      time-unit  & 1d         & 1h         & 10min      & 1d         & 1h         & 10min \\
      \midrule
      \multirow{2}{3cm}{period of time from ... to ...}
                &  15.07.85  &  14.07.09  &  03.01.11  &  02.08.93  &  17.12.99  &  03.01.11\\
                &  25.01.13  &  25.01.13  &  25.01.13  &  25.01.13  &  25.01.13  &  25.01.13 \\
      number of candles
                & \mrA{ 7151}   & \mrA{21964}   & \mrA{77235}   & \mrA{ 4939}   & \mrA{42622}   & \mrA{44329} \\
      number of trends
                & \mrA{   33}   & \mrA{   59}   & \mrA{  299}   & \mrA{   18}   & \mrA{   93}   & \mrA{  286} \\
      \midrule
      $\#($1--2--3$)$
                & \mrA{   44}   & \mrA{   42}   & \mrA{  206}   & \mrA{   16}   & \mrA{   83}   & \mrA{  206} \\
      probability of
                & \mrN{ 0.39}   & \mrN{ 0.50}   & \mrN{ 0.52}   & \mrN{ 0.62}   & \mrN{ 0.55}   & \mrN{ 0.49} \\
      activate a trend&&&&&&\\
      $\mathbb{E}(R_{\textnormal{1--2--3}, \textnormal{ATR}})$
                & \mrA{ 3.53}   & \mrA{ 5.20}   & \mrA{ 4.47}   & \mrA{ 4.24}   & \mrA{ 6.62}   & \mrA{ 3.67} \\
      $\mathbb{E}(G_{\textnormal{1--2--3}, \textnormal{ATR}})$
                & \mrA{ 3.43}   & \mrA{ 4.80}   & \mrA{ 3.87}   & \mrA{ 3.00}   & \mrA{ 5.37}   & \mrA{ 3.28} \\
      $\mathbb{E}(R_{\textnormal{1--2--3}, \%})$
                & \mrA{ 0.58}   & \mrA{ 0.56}   & \mrA{ 0.58}   & \mrA{ 0.62}   & \mrA{ 0.61}   & \mrA{ 0.57} \\
      $\mathbb{E} (\frac{\textnormal{corr. height}}{\textnormal{ini. move. height}})$
                & \mrA{ 0.69}   & \mrA{ 0.76}   & \mrA{ 0.70}   & \mrA{ 0.64}   & \mrA{ 0.73}   & \mrA{ 0.73} \\
      \midrule
      $\#($3--2--3$)$
                & \mrA{   25}   & \mrA{   27}   & \mrA{   95}   & \mrA{   14}   & \mrA{   51}   & \mrA{  103} \\
      probability pass P2
                & \mrN{ 0.60}   & \mrN{ 0.52}   & \mrN{ 0.40}   & \mrN{ 0.50}   & \mrN{ 0.65}   & \mrN{ 0.39} \\
      after a 3--2--3&&&&&&\\
      $\mathbb{E}(R_{\textnormal{3--2--3}, \textnormal{ATR}})$
                & \mrA{ 4.45}   & \mrA{ 5.99}   & \mrA{ 4.36}   & \mrA{ 4.71}   & \mrA{ 6.90}   & \mrA{ 3.63} \\
      $\mathbb{E}(G_{\textnormal{3--2--3}, \textnormal{ATR}})$
                & \mrA{ 3.03}   & \mrA{ 4.01}   & \mrA{ 3.71}   & \mrA{ 4.17}   & \mrA{ 4.80}   & \mrA{ 3.83} \\
      $\mathbb{E}(R_{\textnormal{3--2--3}, \%})$
                & \mrA{ 0.62}   & \mrA{ 0.63}   & \mrA{ 0.57}   & \mrA{ 0.60}   & \mrA{ 0.65}   & \mrA{ 0.54} \\
      $\mathbb{E} (\frac{\textnormal{corr. height}}{\textnormal{move. height}})$
                & \mrA{ 0.68}   & \mrA{ 0.74}   & \mrA{ 0.72}   & \mrA{ 0.74}   & \mrA{ 0.77}   & \mrA{ 0.76} \\
      \midrule
      $\#($2--3--2$)$
                & \mrA{   69}   & \mrA{  104}   & \mrA{  472}   & \mrA{   32}   & \mrA{  187}   & \mrA{  495} \\
      probability pass P2
                & \mrN{ 0.52}   & \mrN{ 0.43}   & \mrN{ 0.37}   & \mrN{ 0.44}   & \mrN{ 0.50}   & \mrN{ 0.42} \\
      after a 2--3--2&&&&&&\\
      $\mathbb{E}$(rel. dur. of break)
                & \mrA{ 2.38}   & \mrA{ 1.35}   & \mrA{ 2.02}   & \mrA{ 1.73}   & \mrA{ 1.30}   & \mrA{ 2.42} \\
      $\mathbb{E}$(rel. dur. of dyn.)
                & \mrA{ 4.39}   & \mrA{ 2.98}   & \mrA{ 4.28}   & \mrA{ 4.51}   & \mrA{ 2.87}   & \mrA{ 4.46} \\
      $\mathbb{E}$(dynamic)
                & \mrA{ 2.18}   & \mrA{ 1.82}   & \mrA{ 1.88}   & \mrA{ 2.15}   & \mrA{ 1.75}   & \mrA{ 2.09} \\
      $\mathbb{E}$(rel. dur. of
                & \mrN{ 5.61}   & \mrN{ 4.32}   & \mrN{ 5.87}   & \mrN{ 5.42}   & \mrN{ 3.71}   & \mrN{ 5.90} \\
      \phantom{$\mathbb{E}($}lag. dyn.)&&&&&&\\
      $\mathbb{E}$(lag. dynamic)
                & \mrA{ 1.36}   & \mrA{ 1.08}   & \mrA{ 0.97}   & \mrA{ 1.36}   & \mrA{ 0.99}   & \mrA{ 1.16} \\
      $\mathbb{E} (\frac{\textnormal{move. height}}{\textnormal{ATR}(\textnormal{P}3)})$
                & \mrA{ 9.88}   & \mrA{12.22}   & \mrA{ 9.66}   & \mrA{11.83}   & \mrA{13.66}   & \mrA{ 8.34} \\
      $\mathbb{E} (\frac{\textnormal{corr. height}}{\textnormal{ATR}(\textnormal{P}3)})$
                & \mrA{ 4.73}   & \mrA{ 7.03}   & \mrA{ 5.47}   & \mrA{ 5.75}   & \mrA{ 7.85}   & \mrA{ 4.43} \\
      $\mathbb{E} (\frac{\textnormal{move. height}}{\textnormal{low}(\textnormal{P}3)})$
                & \mrA{0.0873}  & \mrA{0.0242}  & \mrA{0.0079}  & \mrA{0.2311}  & \mrA{0.0778}  & \mrA{0.0179} \\
      $\mathbb{E} (\frac{\textnormal{corr. height}}{\textnormal{low}(\textnormal{P}3)})$
                & \mrA{0.0437}  & \mrA{0.0141}  & \mrA{0.0045}  & \mrA{0.1186}  & \mrA{0.0455}  & \mrA{0.0096} \\
      \midrule
      $\mathbb{E}$(number of
                & \mrN{ 3.09}   & \mrN{ 2.76}   & \mrN{ 2.58}   & \mrN{ 2.78}   & \mrN{ 3.01}   & \mrN{ 2.73} \\
      \phantom{$\mathbb{E}($}movements)&&&&&&\\
      \bottomrule
    \end{tabular}
  }
  \label{tab:other_critera_eurusd_fdax}
\end{table}

\begin{table}[p] %tab_3.2
  \centering
  \caption{Some quality criteria for Gold and Crude Oil with different aggregations.}
  {\footnotesize
    \newcommand{\mc}[3]{\multicolumn{#1}{#2}{#3}}
    \newcommand{\mr}[3]{\multirow{#1}{#2}{#3}}
    \newcommand{\mrN}[1]{\multirow{2}{*}{#1}}
    \newcommand{\mrA}[1]{\multirow{1}{*}{#1}}
    \begin{tabular}{lcccccc}
      \toprule
      underlying &\mc{3}{c}{Gold} &\mc{3}{c}{Crude Oil}\\
      \cmidrule(lr){2-4} \cmidrule(l){5-7}
      time-unit  & 1d         & 1h         & 10min      & 1d         & 1h         & 10min \\
      \midrule
      \multirow{2}{3cm}{period of time from ... to ...}
                &  14.09.90  &  11.07.05  &  03.01.11  &  14.09.90  &  29.11.04  &  03.01.11\\
                &  25.01.13  &  25.01.13  &  25.01.13  &  25.01.13  &  25.01.13  &  25.01.13 \\
      number of candles
                & \mrA{ 5632}   & \mrA{42901}   & \mrA{74356}   & \mrA{ 5622}   & \mrA{44907}   & \mrA{74310} \\
      number of trends
                & \mrA{   33}   & \mrA{  136}   & \mrA{  258}   & \mrA{   20}   & \mrA{  170}   & \mrA{  458} \\
      \midrule
      $\#($1--2--3$)$
                & \mrA{   44}   & \mrA{  122}   & \mrA{  198}   & \mrA{   25}   & \mrA{  142}   & \mrA{  453} \\
      probability of
                & \mrN{ 0.43}   & \mrN{ 0.46}   & \mrN{ 0.42}   & \mrN{ 0.28}   & \mrN{ 0.49}   & \mrN{ 0.48} \\
      activate a trend&&&&&&\\
      $\mathbb{E}(R_{\textnormal{1--2--3}, \textnormal{ATR}})$
                & \mrA{ 3.56}   & \mrA{ 4.94}   & \mrA{ 4.36}   & \mrA{ 4.52}   & \mrA{ 5.39}   & \mrA{ 3.23} \\
      $\mathbb{E}(G_{\textnormal{1--2--3}, \textnormal{ATR}})$
                & \mrA{ 3.66}   & \mrA{ 5.15}   & \mrA{ 4.25}   & \mrA{ 3.79}   & \mrA{ 4.16}   & \mrA{ 3.47} \\
      $\mathbb{E}(R_{\textnormal{1--2--3}, \%})$
                & \mrA{ 0.53}   & \mrA{ 0.53}   & \mrA{ 0.56}   & \mrA{ 0.59}   & \mrA{ 0.60}   & \mrA{ 0.54} \\
      $\mathbb{E} (\frac{\textnormal{corr. height}}{\textnormal{ini. move. height}})$
                & \mrA{ 0.71}   & \mrA{ 0.72}   & \mrA{ 0.68}   & \mrA{ 0.71}   & \mrA{ 0.75}   & \mrA{ 0.66} \\
      \midrule
      $\#($3--2--3$)$
                & \mrA{   20}   & \mrA{   61}   & \mrA{  114}   & \mrA{   12}   & \mrA{   73}   & \mrA{  163} \\
      probability pass P2
                & \mrN{ 0.40}   & \mrN{ 0.43}   & \mrN{ 0.44}   & \mrN{ 0.58}   & \mrN{ 0.44}   & \mrN{ 0.28} \\
      after a 3--2--3&&&&&&\\
      $\mathbb{E}(R_{\textnormal{3--2--3}, \textnormal{ATR}})$
                & \mrA{ 4.50}   & \mrA{ 5.08}   & \mrA{ 4.93}   & \mrA{ 4.55}   & \mrA{ 5.32}   & \mrA{ 3.52} \\
      $\mathbb{E}(G_{\textnormal{3--2--3}, \textnormal{ATR}})$
                & \mrA{ 3.42}   & \mrA{ 5.37}   & \mrA{ 4.82}   & \mrA{ 3.65}   & \mrA{ 4.28}   & \mrA{ 4.39} \\
      $\mathbb{E}(R_{\textnormal{3--2--3}, \%})$
                & \mrA{ 0.60}   & \mrA{ 0.54}   & \mrA{ 0.55}   & \mrA{ 0.60}   & \mrA{ 0.61}   & \mrA{ 0.49} \\
      $\mathbb{E} (\frac{\textnormal{corr. height}}{\textnormal{move. height}})$
                & \mrA{ 0.75}   & \mrA{ 0.75}   & \mrA{ 0.69}   & \mrA{ 0.76}   & \mrA{ 0.73}   & \mrA{ 0.67} \\
      \midrule
      $\#($2--3--2$)$
                & \mrA{   61}   & \mrA{  252}   & \mrA{  435}   & \mrA{   38}   & \mrA{  302}   & \mrA{  720} \\
      probability pass P2
                & \mrN{ 0.46}   & \mrN{ 0.46}   & \mrN{ 0.41}   & \mrN{ 0.47}   & \mrN{ 0.44}   & \mrN{ 0.36} \\
      after a 2--3--2&&&&&&\\
      $\mathbb{E}$(rel. dur. of break)
                & \mrA{ 1.52}   & \mrA{ 2.54}   & \mrA{ 3.60}   & \mrA{ 1.66}   & \mrA{ 2.90}   & \mrA{ 2.97} \\
      $\mathbb{E}$(rel. dur. of dyn.)
                & \mrA{ 3.21}   & \mrA{ 4.21}   & \mrA{ 5.35}   & \mrA{ 3.14}   & \mrA{ 5.27}   & \mrA{ 4.59} \\
      $\mathbb{E}$(dynamic)
                & \mrA{ 2.07}   & \mrA{ 1.86}   & \mrA{ 2.02}   & \mrA{ 1.81}   & \mrA{ 1.77}   & \mrA{ 2.06} \\
      $\mathbb{E}$(rel. dur. of
                & \mrN{ 4.41}   & \mrN{ 5.63}   & \mrN{ 7.12}   & \mrN{ 4.33}   & \mrN{ 6.69}   & \mrN{ 6.05} \\
      \phantom{$\mathbb{E}($}lag. dyn.)&&&&&&\\
      $\mathbb{E}$(lag. dynamic)
                & \mrA{ 1.15}   & \mrA{ 1.09}   & \mrA{ 1.06}   & \mrA{ 0.90}   & \mrA{ 0.98}   & \mrA{ 1.09} \\
      $\mathbb{E} (\frac{\textnormal{move. height}}{\textnormal{ATR}(\textnormal{P}3)})$
                & \mrA{ 9.71}   & \mrA{11.96}   & \mrA{11.63}   & \mrA{10.34}   & \mrA{10.81}   & \mrA{ 9.03} \\
      $\mathbb{E} (\frac{\textnormal{corr. height}}{\textnormal{ATR}(\textnormal{P}3)})$
                & \mrA{ 5.23}   & \mrA{ 6.63}   & \mrA{ 6.15}   & \mrA{ 5.77}   & \mrA{ 6.35}   & \mrA{ 4.66} \\
      $\mathbb{E} (\frac{\textnormal{move. height}}{\textnormal{low}(\textnormal{P}3)})$
                & \mrA{0.1209}  & \mrA{0.0431}  & \mrA{0.0143}  & \mrA{0.3026}  & \mrA{0.0655}  & \mrA{0.0173} \\
      $\mathbb{E} (\frac{\textnormal{corr. height}}{\textnormal{low}(\textnormal{P}3)})$
                & \mrA{0.0649}  & \mrA{0.0240}  & \mrA{0.0076}  & \mrA{0.1733}  & \mrA{0.0383}  & \mrA{0.0091} \\
      \midrule
      $\mathbb{E}$(number of
                & \mrN{ 2.85}   & \mrN{ 2.85}   & \mrN{ 2.69}   & \mrN{ 2.90}   & \mrN{ 2.78}   & \mrN{ 2.57} \\
      \phantom{$\mathbb{E}($}movements)&&&&&&\\
      \bottomrule
    \end{tabular}
  }
  \label{tab:other_critera_gold_oil}
\end{table}

%%%

\subsection{Position of the new P2 for 2--3--2}\label{subsec:pos_p2} % 3.3

 In the last subsection we only used the (empirical) expectation to study the arithmetic mean of the observed quantities. Of course the expectation by itself
 does not tell us the whole truth. Values can have large fluctuations so that the expectation is less meaningful. However, we can not give a full discussion
 on all quantifies of Table~\ref{tab:other_critera_eurusd_fdax} and Table~\ref{tab:other_critera_gold_oil}. Hence we will concentrate on one of the most
 important quantities which is the dynamic of 2--3--2 situations.

\begin{figure}[!t] %fig_3.4
  \centering
  \includegraphics[width=0.7\linewidth]{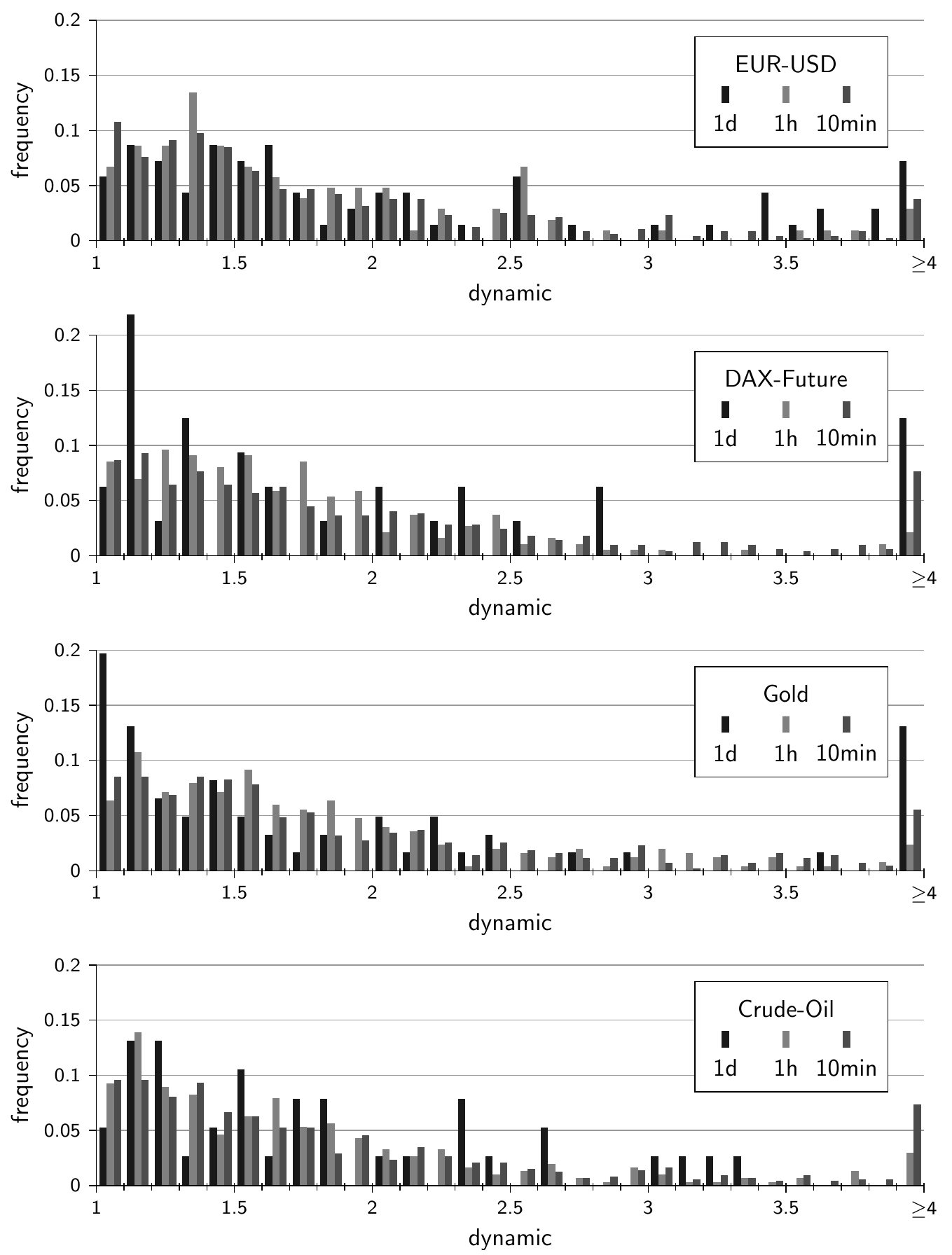}
  \caption{Frequency distribution of the dynamics, i.e. the relative positions of $\textnormal{P2}_{\textnormal{new}}$ in a situation of
  type 2--3--2 of Definition~\ref{def:trend_properties}.}
  \label{fig:positions_of_p2}
\end{figure}

 In Figure~\ref{fig:positions_of_p2} one sees the relative frequency distribution of the dynamic for all four underlyings. We can see, that the dynamic is
 concentrated between a value of 1 and 2.5 and there are even values which are greater of equal to 4. On 1 day basis there is a noticeable large number of
 situations with a dynamic of 4 and larger on the EUR-USD, DAX-Future and Gold while there is none for the Crude Oil. This tells us that especially on
 DAX-Future and Gold stable long term trends are possible. The frequency distributions for 1h and 10min look more smooth. Here on 10min basis the dynamic
 also can be greater or equal to 4 with a probability of at least 5\%, while the probability for 1h time aggregation is always below 5\%. This is the most
 significant difference between the distribution of different time aggregations, which of course effects the expectation of the dynamic,
 see Tables~\ref{tab:other_critera_eurusd_fdax} and~\ref{tab:other_critera_gold_oil}. In Gold on day basis the number of 2--3--2 situations with very small
 dynamics is extremely large. A lot of movements are killed right away which could be caused by possible market manipulations. All distributions in
 Figure~\ref{fig:positions_of_p2} look approximately like Log-normal distributions.

 Besides the dynamic the relative duration of the dynamic, i.e. the relative duration of the movement phase, is needed to identify the position of
 $\textnormal{P2}_{\textnormal{new}}$ in 2--3--2 Situations. Since a large dynamic is preferable we are interested in the probability of exceeding a
 given value. We therefore do not use the cumulative distribution function in the classical sense, but we use the probability of a position of the
 new P2 which has a dynamic and a relative duration of dynamic larger than given values, i.e. we use
 \begin{align}\label{eq:cum_dist_func} %NR_3.1
   \tilde{F}(x,y) := \mathbb{P}\left(
    \textnormal{rel. duration of dyn.}\geq x,\,\,
    \textnormal{dynamic}\geq y
  \right)
 \end{align}
 for $x\geq 0$ and $y\geq 1$. From Tables~\ref{tab:other_critera_eurusd_fdax} and~\ref{tab:other_critera_gold_oil} we directly get the expectation of
 this two dimensional distribution, which is defined by the expectations of each component, i.e.
 \begin{align}\label{eq:multi_exp} %NR_3.2
   \mathbb{E}\left(
     \left[\begin{matrix}
       \textnormal{rel. duration of dyn.}\\
       \textnormal{dynamic}
     \end{matrix}\right]
   \right)
    &:= \left[\begin{matrix}
       \mathbb{E}(\textnormal{rel. duration of dyn.})\\
       \mathbb{E}(\textnormal{dynamic})
     \end{matrix}\right],
 \end{align}
 see e.g. \cite[Definition 3.3]{Mittelhammer1999}. Analogously one can define the expectations for $\textnormal{P2}_{\textnormal{break}}$ and the
 identification of $\textnormal{P2}_{\textnormal{new}}$ by
 \begin{align}\label{eq:multi_exp2} %NR_3.3
   \mathbb{E}\left(
     \left[\begin{matrix}
       \textnormal{rel. dur. of break}\\
       \textnormal{``dynamic of break''}
     \end{matrix}\right]
   \right)
    &:= \left[\begin{matrix}
       \mathbb{E}(\textnormal{rel. dur. of break})\\
       1
     \end{matrix}\right]\\
   \label{eq:multi_exp3} %NR_3.4
   \text{and}\qquad
   \mathbb{E}\left(
     \left[\begin{matrix}
       \textnormal{rel. dur. of lag. dyn.}\\
       \textnormal{lag. dynamic}
     \end{matrix}\right]
   \right)
    &:= \left[\begin{matrix}
       \mathbb{E}(\textnormal{rel. dur. of lag. dyn.})\\
       \mathbb{E}(\textnormal{lag. dynamic})
     \end{matrix}\right].
 \end{align}

\begin{figure}
  \def\WIDTH{0.45\linewidth}
  \def\HSPACE{-2mm}
  \def\VSPACE{-3mm}
  \centering
  \subfloat[EUR-USD]{
    \parbox{\WIDTH}{
      \includegraphics[width=\linewidth]{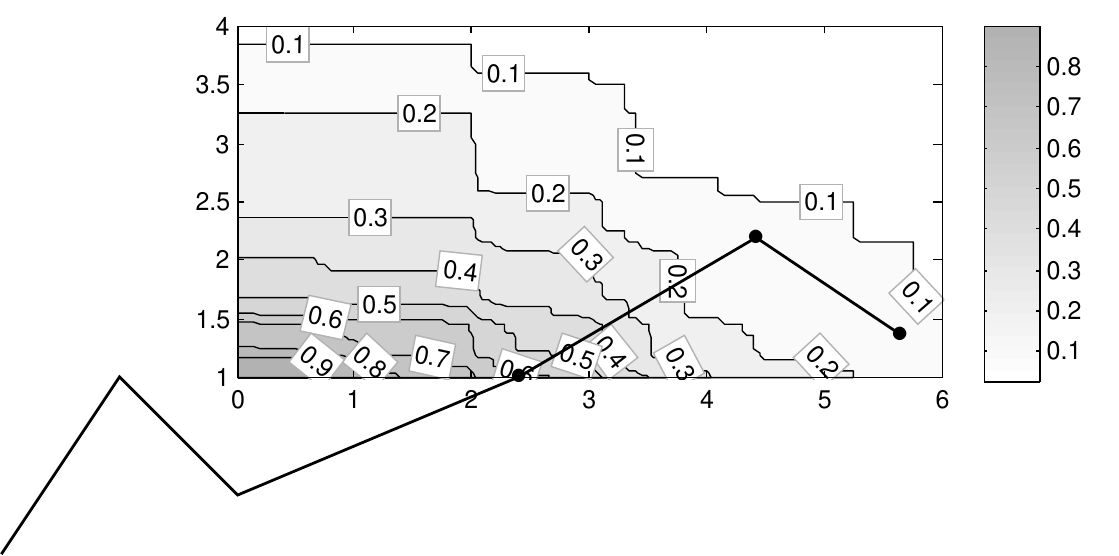}
      \\
      \includegraphics[width=\linewidth]{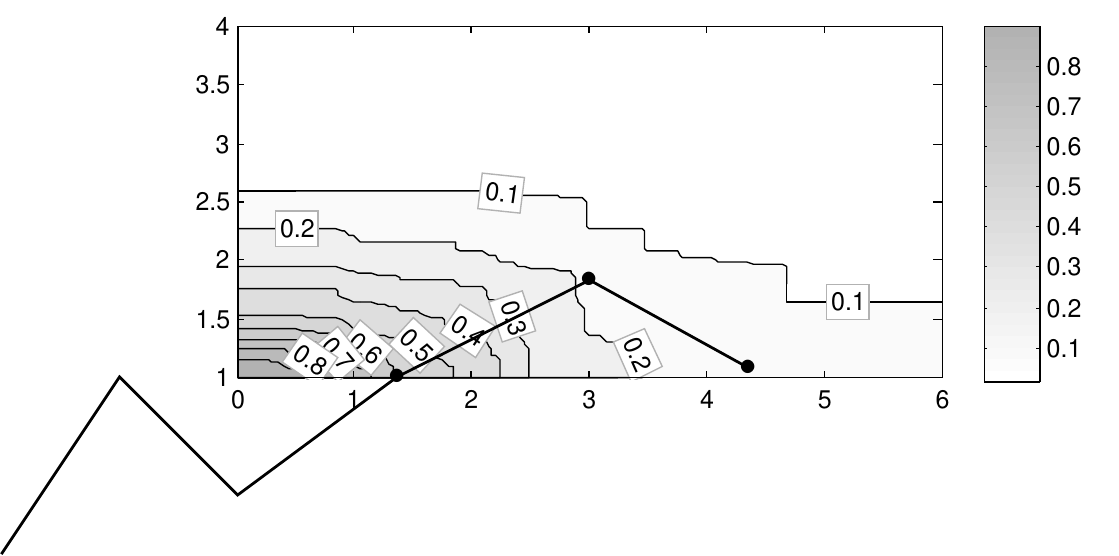}
      \\
      \includegraphics[width=\linewidth]{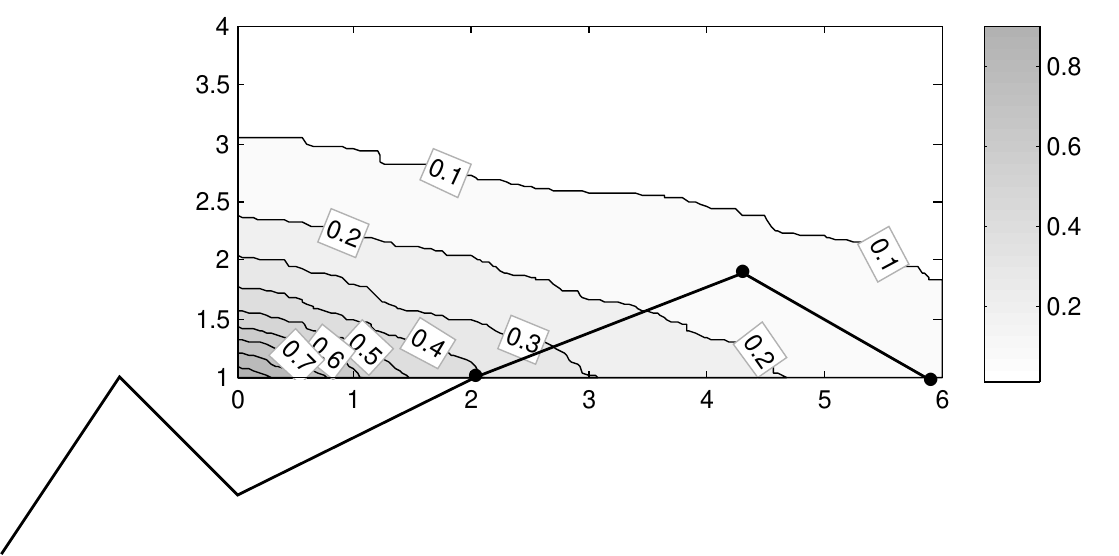}
    }
    \label{fig:eur_usd_distribution}
  }
  \subfloat[DAX-Future]{
    \parbox{\WIDTH}{
      \includegraphics[width=\linewidth]{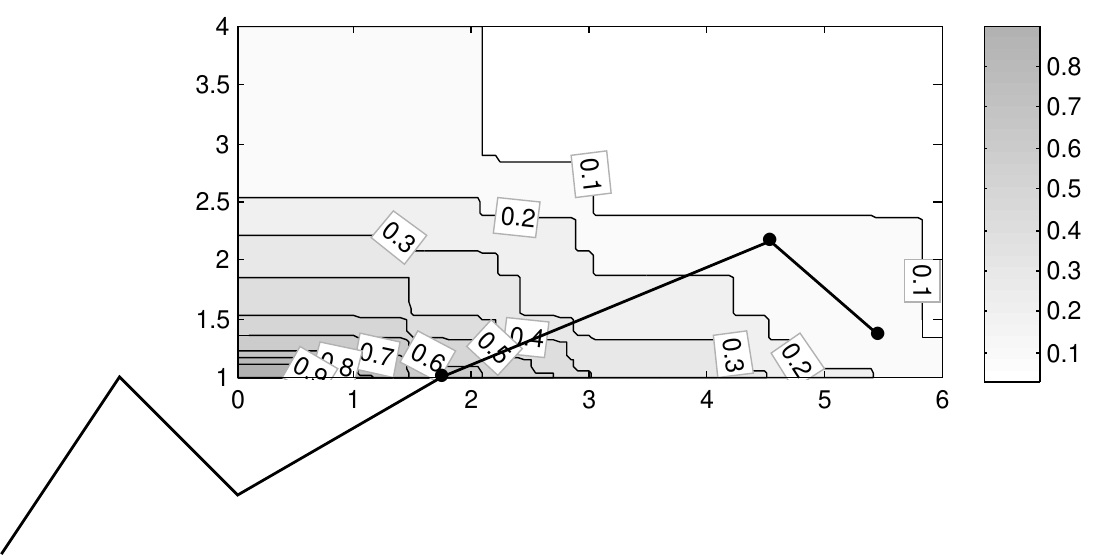}
      \\
      \includegraphics[width=\linewidth]{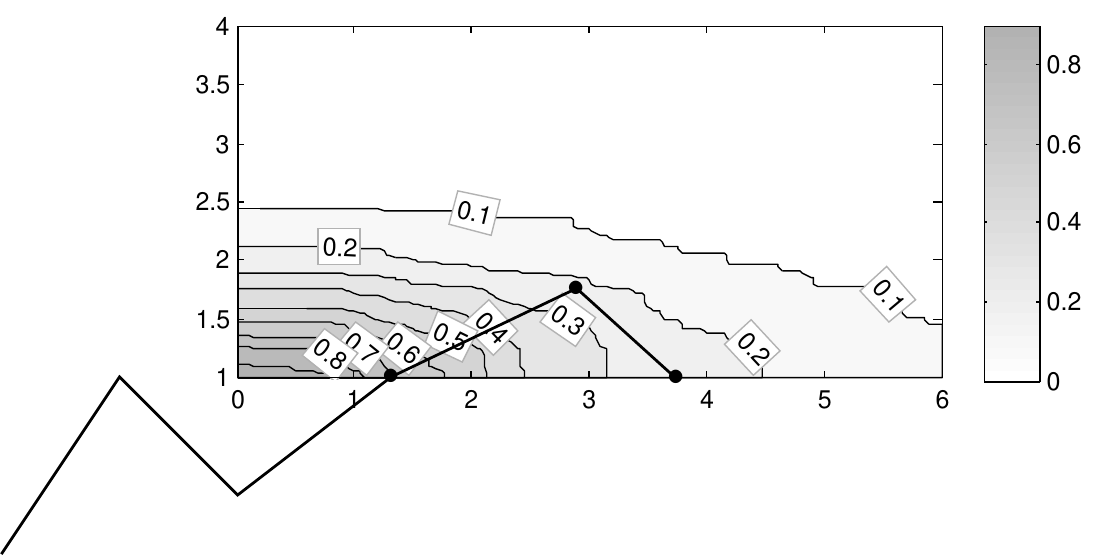}
      \\
      \includegraphics[width=\linewidth]{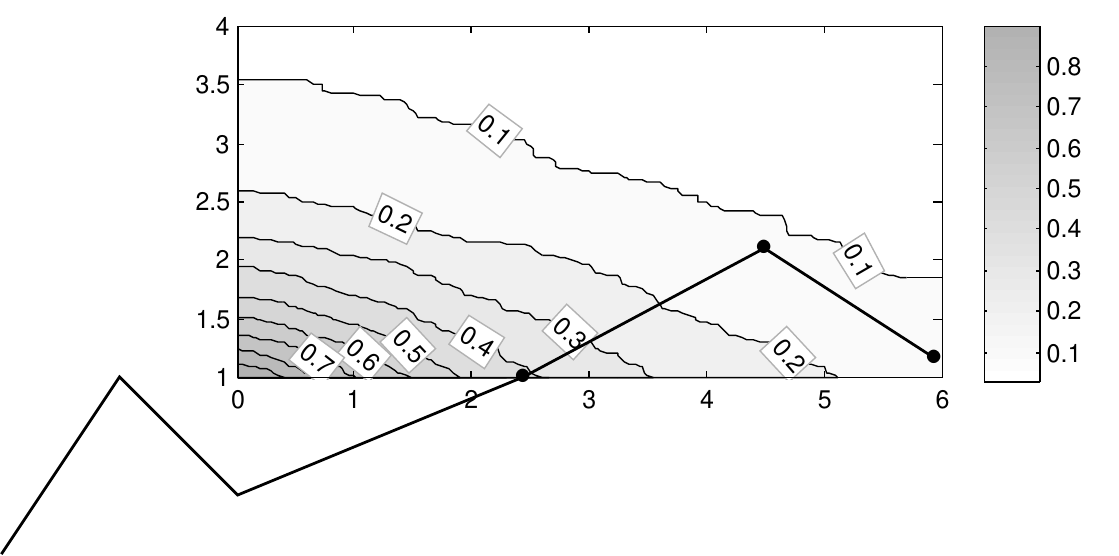}
    }
    \label{fig:dax_distribution}
  }
  \caption{The reversed cumulative distribution function $\tilde{F}(x,y)$ from \eqref{eq:cum_dist_func} is plotted and the expectations
  from~\eqref{eq:multi_exp},~\eqref{eq:multi_exp2} and ~\eqref{eq:multi_exp3} are marked.
  The aggregation is 1 day, 1 hour and 10 minutes from top to bottom.}
  \label{fig:eur_usd_dax_distribution}
\end{figure}

\begin{figure}[t]
  \def\WIDTH{0.45\linewidth}
  \def\HSPACE{-2mm}
  \def\VSPACE{-3mm}
  \centering
  \subfloat[Gold]{
    \parbox{\WIDTH}{
      \includegraphics[width=\linewidth]{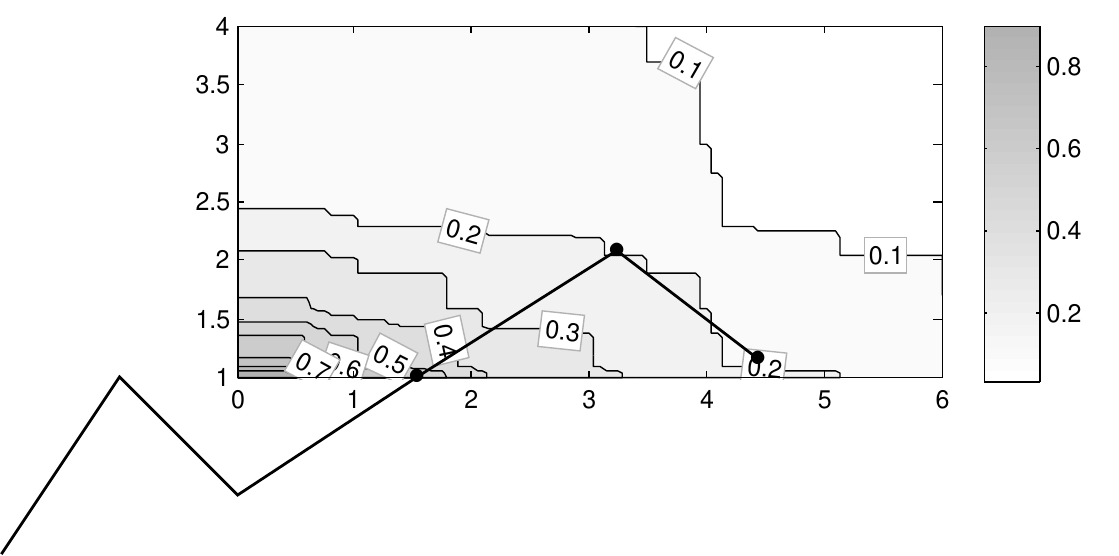}
      \\
      \includegraphics[width=\linewidth]{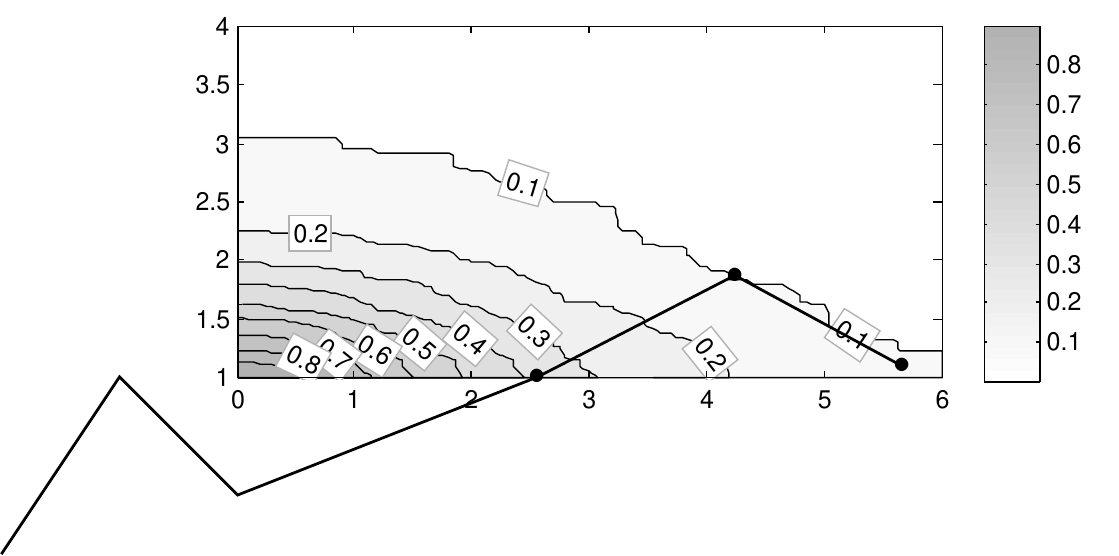}
      \\
      \includegraphics[width=\linewidth]{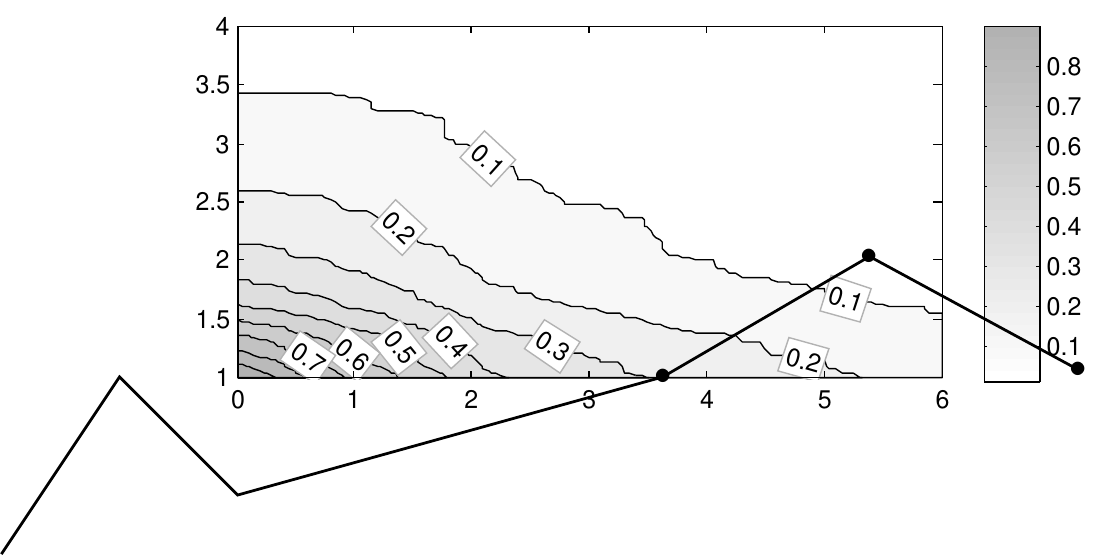}
    }
    \label{fig:gold_distribution}
  }
  \subfloat[Crude Oil]{
    \parbox{\WIDTH}{
      \includegraphics[width=\linewidth]{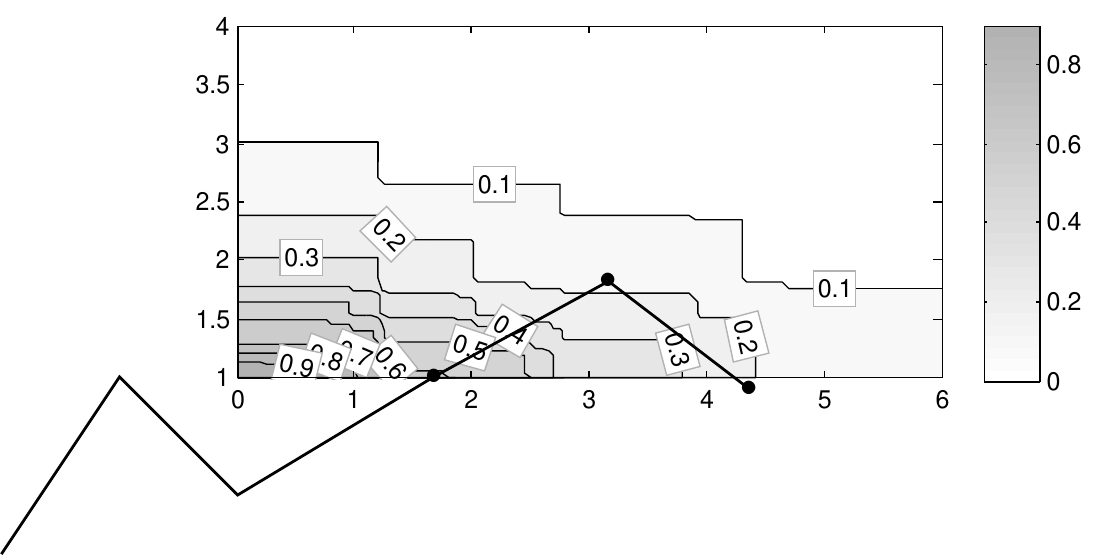}
      \\
      \includegraphics[width=\linewidth]{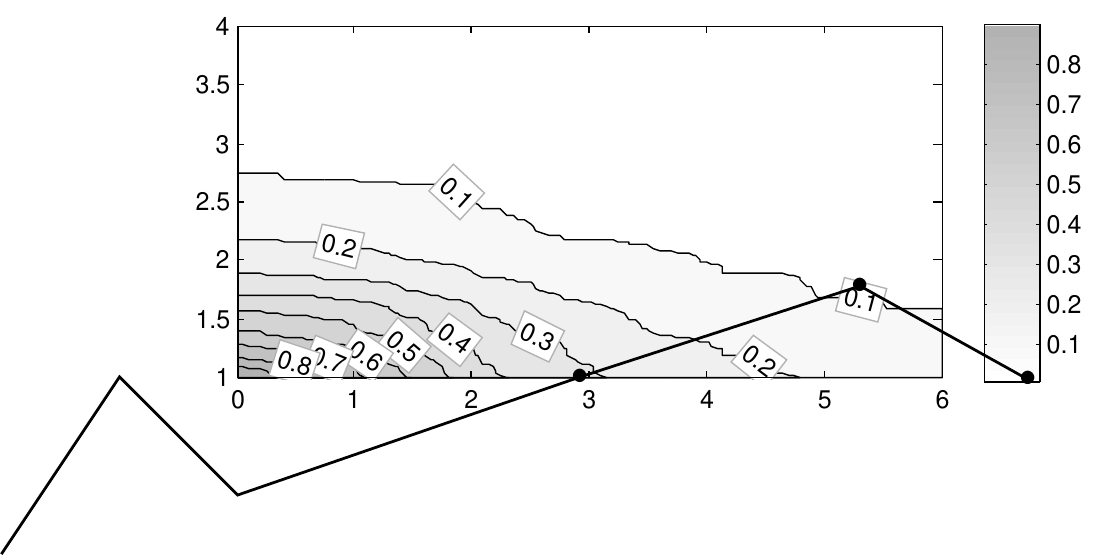}
      \\
      \includegraphics[width=\linewidth]{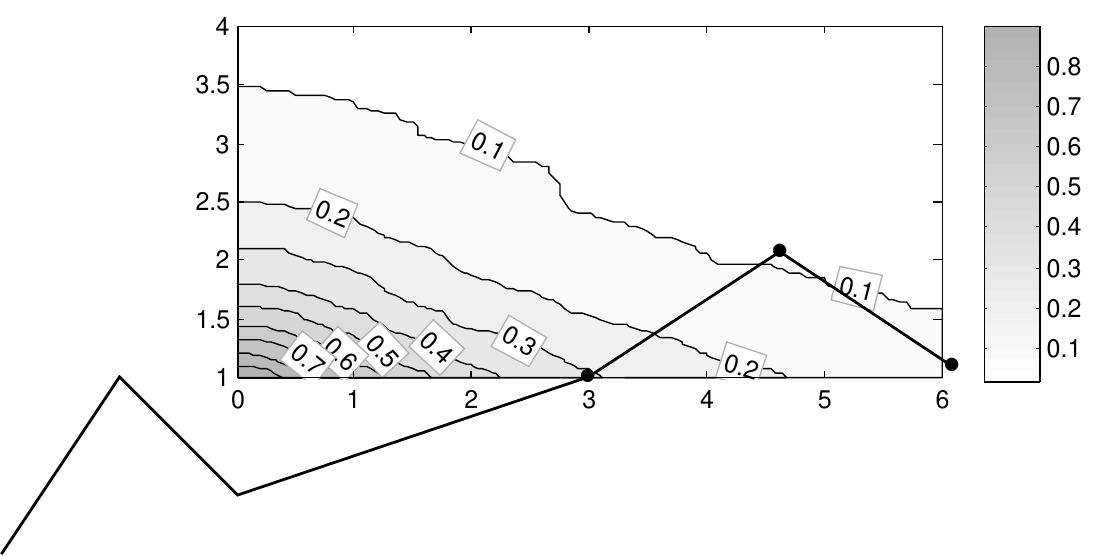}
    }
    \label{fig:oil_distribution}
  }
  \caption{The reversed cumulative distribution function $\tilde{F}(x,y)$ from \eqref{eq:cum_dist_func} is plotted and the expectations
  from~\eqref{eq:multi_exp},~\eqref{eq:multi_exp2} and ~\eqref{eq:multi_exp3} are marked.
  The aggregation is 1 day, 1 hour and 10 minutes from top to bottom.}
  \label{fig:gold_oil_distribution}
\end{figure}

 In Figures~\ref{fig:eur_usd_dax_distribution} and \ref{fig:gold_oil_distribution} the reversed cumulative distribution function from \eqref{eq:cum_dist_func}
 is plotted for the EUR-USD, DAX-Future, Gold and Crude Oil, respectively, for all three time aggregations. The shape looks quite similar for all these
 functions. One clearly sees that the probability of a long duration of the movement phase is much higher than the probability of a high move.
 Another observation is that the velocities of the price value on average are roughly the same for the move from
 $\textnormal{P3}$ to $\textnormal{P2}_{\textnormal{break}}$ and from $\textnormal{P2}_{\textnormal{break}}$ to $\textnormal{P2}_{\textnormal{new}}$.
 This means that the whole movement is slow compared to the correction phase.

%%%%%%%%%%%%%%%%%%%%

\section{Conclusions}

 To identify a trend the strict trend definition from Dow has been used. The relevant extrema of the chart were found by the MinMax algorithm of
 \cite{Maier-Paape2013} with the MACD as SAR process. Except for the timescale we always used the default parameter settings and did not need any
 optimization. We only set the timescale appropriate to match the dominant wavelength, which we obtained from the time shift with the highest
 autocorrelation of the price chart. We then saw many different results for trends on four different underlyings. One important observation is that
 the movement phase is relatively slow in comparison with the correction. We also see some side effects from the efficient-market hypothesis:
 We saw that the gain $G_{\textnormal{1--2--3,\%}}$ and $G_{\textnormal{3--2--3,\%}}$ is roughly at 50\% and the probability of getting this gain, i.e.
 of reaching the last P2, is also about 50\%. Improvements of these results could be appropriate selections of high potential situations. This could be
 done by filters and the use of nested trends.

\newpage

% Literaturverzeichnis
\bibliographystyle{gerplain}
\bibliography{finance}

\begin{thebibliography}{10}

\bibitem{Appel2005}
{\scshape Appel, Gerald}: {\em Technical Analysis: Power Tools for Active
  Investors}.
\newblock Financial Times Prentice Hall, Upper Saddle River, New Jersey, 2005.

\bibitem{Cene2011}
{\scshape Cene, E.}: {\em Professioneller B\"orsenhandel}.
\newblock FinanzBuch Verlag, M\"unchen, 2011.

\bibitem{CK1994}
{\scshape Chande, T.~S.} \btxandlong{}\ {\scshape S.~Kroll}: {\em The New Technical
  Trader}.
\newblock John Wiley \& Sons, Hoboken, 1994.

\bibitem{Duerschner2013}
{\scshape D\"urschner, M.~G.}: {\em Technische Analyse mit EMD}.
\newblock Wiley, Hoboken, New Jersey, 2013.

\bibitem{FB1966}
{\scshape Fama, E.~F.} \btxandlong{}\ {\scshape M.~E. Blume}: {\em Filter Rules and
  Stock-Market Trading}.
\newblock Journal of Business, 39:226--241, 1966.

\bibitem{Heckmann2009}
{\scshape Heckmann, T.}: {\em Markttechnische Handelssysteme, quantitative
  Kursmuster und saisonale Kursanomalien}.
\newblock Eul Verlag, Lohmar, 2009.

\bibitem{HSL+1998}
{\scshape Huang, N.~E.}, {\scshape Z.~Shen}, {\scshape S.~R. Long}, {\scshape M.~C. Wu}, {\scshape H.~H.
  Shih}, {\scshape Q.~Zheng}, {\scshape N.-C. Yen}, {\scshape C.~C. Tung} \btxandlong{}\ {\scshape
  H.~H. Liu}: {\em The empirical mode decomposition and the Hilbert spectrum
  for nonlinear and non-stationary time series analysis}.
\newblock Proceedings of the Royal Society of London, 454(1971):903--995, 1998.

\bibitem{Leuthold1972}
{\scshape Leuthold, R.~M.}: {\em Random walk and price trends: The live cattle
  futures market}.
\newblock The Journal of Finance, 27(4):879--889, 1972.

\bibitem{Maier-Paape2013}
{\scshape Maier-Paape, S.}: {\em Automatic One Two Three}.
\newblock Quantitative Finance, 2013.

\bibitem{MV2006}
{\scshape Meyberg, K.} \btxandlong{}\ {\scshape P.~Vachenauer}: {\em H\"ohere Mathematik
  2}.
\newblock Springer, Heidelberg, 2006.

\bibitem{Mittelhammer1999}
{\scshape Mittelhammer, R.~C.}: {\em Mathematical Statistics for Economics and
  Business}.
\newblock Springer, Heidelberg, 1999.

\bibitem{Poulos1991}
{\scshape Poulos, E.~M.}: {\em Of Trends and Random Walks}.
\newblock Technical analysis of Stocks \& Commodities, 9(2):49--52, 1991.

\bibitem{Sidney1961}
{\scshape Sidney, S.~A.}: {\em Price Movements in speculative Markets: Trends or
  Random Walks}.
\newblock Industrial Management Review Industrial Management Review, 2:7--26,
  1961.

\bibitem{Voigt2010}
{\scshape Voigt, M.}: {\em Das gro{\ss}e Buch der Markttechnik}.
\newblock FinanzBuch Verlag, M\"unchen, 7. \btxeditionlong{}, 2010.

\bibitem{Wilder1978}
{\scshape Wilder, W.~J.}: {\em New Concepts in Technical Trading Systems}.
\newblock Trend Research, McLeansville, North Carolina, 1978.

\end{thebibliography}

\end{document}